\documentclass[a4paper,12pt]{article}
\pdfoutput=1
\usepackage[intlimits,centertags]{amsmath}
\usepackage{amssymb,amsfonts}
\usepackage{bbm}
\usepackage{mathrsfs}
\usepackage{cite}
\usepackage[scale={0.78,0.8}]{geometry}
\usepackage{multirow}
\usepackage{relsize}
\usepackage[subrefformat=parens,labelformat=parens]{subfig}
\usepackage{xcolor}
\usepackage{caption}
\usepackage{cancel}
\usepackage{graphics,graphicx}
\usepackage{float}
\numberwithin{equation}{section}

\def\be{\begin{equation}}
\def\ee{\end{equation}}
\def\beq{\begin{align}}
\def\eeq{\end{align}}
\def\beqa{\begin{eqnarray}}
\def\eeqa{\end{eqnarray}}

\begin{document}

\pagestyle{empty}
\rightline{DESY 19-160}
\vspace{1.2cm}

\vskip 1.5cm

\begin{center}
\LARGE{Flattened Axion Monodromy Beyond Two Derivatives}
\\[13mm]
\large{Francisco G. Pedro$^{1,2}$, Alexander Westphal$^{3}$ } \\[6mm]
\small{${}^1$ Dipartimento di Fisica e Astronomia, Universit\`a di Bologna, \\ via Irnerio 46, 40126 Bologna, Italy\\
${}^2$INFN, Sezione di Bologna, viale Berti Pichat 6/2, 40127 Bologna, Italy\\ [6mm]}
\small{
${}^3$Deutsches Elektronen-Synchrotron DESY, \\ Theory Group, D-22603 Hamburg, Germany}
\end{center}

\vspace{2cm}
\begin{abstract}

We study string inspired two-field models of large-field inflation based on axion monodromy in the presence of an interacting heavier modulus. This class of models has enough structure to approximate at least part of the backreaction effects known in full string theory, such as kinetic mixing with the axion, and flattening of the scalar potential.  Yet, it is simple enough to fully describe the structure of higher-point curvature perturbation interactions driven by the adjusting modulus backreaction dynamics. We find that the presence of the heavy modulus can be described via two equivalent effective field theories, both of which can incorporate reductions of the speed of sound. Hence, the presence of heavier moduli in axion monodromy inflation constructions will necessarily generate some amount of non-Gaussianity accompanied by changes to $n_s$ and $r$ beyond what results from just from the well known adiabatic flattening backreaction.
\end{abstract}
\newpage

\setcounter{page}{1}
\pagestyle{plain}
\renewcommand{\thefootnote}{\arabic{footnote}}
\setcounter{footnote}{0}

\tableofcontents
%
\section{Introduction}

The scenario of cosmological inflation has become the standard paradigm for generating the initial conditions of the hot FRW `big bang' epoch starting with reheating and sourcing the seeds of structure formation from a nearly scale-invariant spectrum of phase-coherent quantum curvature perturbations. Morever, inflation sources a nearly scale-invariant spectrum of primordial gravitational waves,  yet to be detected. Observational data as obtained e.g. by recent CMB measurements \cite{Akrami:2018odb} provide strong if not yet unequivocal support. However, observations at this stage provide access to just three observables of the inflationary primordial fluctuation power spectra -- the amplitude of the of the curvature perturbations $\mathcal{A}_s$, their spectral tilt $n_s$, and so far an upper limit on the tensor-to-scalar ratio $r$, the fractional power in primordial gravitational waves \cite{Akrami:2018odb}.

Once inflation arises in setups with more than just the minimal ingredients of a single scalar field with a suitably flat scalar potential and a two-derivative kinetic term, generically higher-point interactions between the inflationary perturbations arise, producing non-Gaussianity of varying wave-vector space `shapes' and of magnitude $f_{NL}$. Data has so far only provided upper limits for these parameters\cite{Akrami:2019izv}. A measurement here would greatly constrain the space of allowed models by directly accessing the interaction structure of the inflationary perturbations \cite{Chen:2006nt}. Indeed, measuring non-Gaussianity amounts to detecting  3-point interactions of the curvature perturbation in 'cosmological collider'~\cite{Arkani-Hamed:2015bza,Arkani-Hamed:2018kmz} analogy of lab-based scattering experiments involving interactions between in-going particle states.

Looking at the structure of inflation models encountered in UV completions like string theory is instructive for restricting choices among the very large class of bottom-up non-minimal inflation setups. In particular, higher-point interactions of the inflationary perturbations seem to arise in string theory models of inflation mostly in two ways: either non-Gaussianity originates from the direct fundamental presence of higher-derivative kinetic terms such as in DBI inflation \cite{Alishahiha:2004eh}, or it is due to the presence of more or less heavy 'spectator' moduli fields backreacting to the dynamics of the inflaton scalar (which may be a modulus itself, or alternatively a stringy axion arising from higher-dimensional $p$-form gauge fields).

Given this context, we will focus on toy models of large-field inflation from string theory based on axion monodromy in the presence of an interacting heavier modulus. This class of models will serve to  approximate at least part of the known effects such as flattening of the scalar potential arising the full backreaction of the typically several moduli and/or adjustment of the internal geometry of the six extra dimensions of string theory. At the same time, as large-field models these models maximize the number of potential CMB observables by generating a measurably large primordial tensor mode signal. Yet, this class of setups is simple enough to fully describe the structure of higher-point curvature perturbation interactions driven by the adjusting modulus backreaction dynamics. We will find, that the 2-field effective field theory (EFT) treatment allows us to justify an alternative description in terms of a purely single-field EFT involving a series of higher-derivative (HD) kinetic term corrections dictated by the backreaction dynamics of the modulus we integrated out. These HD terms are known to lead to a reduction of the propagation speed for the scalar perturbations, $c_s$, a parameter that influences both the two and three point curvature correlators. It is worth noting that the current observational bounds on $f_{NL}$ \cite{Akrami:2019izv} allow ample space for subluminal propagation speeds for the scalar perturbations, given the parametric relation $f_{NL}\sim c_s^{-2}$ \cite{Cheung:2007st,Baumann:2011su}. On the other hand the validity of the perturbative treatment also constrains large reductions of the speed of sound  \cite{Leblond:2008gg}. In what follows we will find that such UV motivated axion monodromy inflation models with backreaction from moduli separated only by a finite mass hierarchy \cite{Baumann:2011su} will, in certain regions of parameter space, generate some amount of non-Gaussianity  accompanied by changes to $n_s$ and $r$. The results presented below go beyond the mere the adiabatic flattening backreaction in the scalar potential, widely studied in this context (see eg. \cite{Dong:2010in,McAllister:2014mpa}, with the 4D effective 4-form description of flattened monodromy given in e.g.~\cite{Kaloper:2011jz,DAmico:2017cda}).

We begin in section~\ref{sec:EFTs} by describing the two alternative EFTs to analyse the dynamics and observational signatures of these monodromy models. In section~\ref{sec:KineticMix} we review the effects of pure kinetic modulus-axion mixing in the language of the two EFTs. Section~\ref{sec:PotKinMix} analyses semi-realistic toy models where the modulus and the axion mix both in the kinetic term and via backreaction of the potential. In both sections we restrict to two classes of kinetic couplings, one of them monomial and the other exponential in the canonically normalised modulus field. This is motivated by the fact, that typically in string theory setups volume deformation modes of the extra dimensions are the lightest moduli in the spectrum, and the coupling of these volume moduli to string axions fall into the two classes of kinetic couplings we consider (at least in the case of tree-level supersymmetric Calabi-Yau compactifications). We find that depending on the relative importance of kinetic coupling vs coupling in the potential, the simple single-field predictions of axion monodromy inflation $n_s-1=-(p+2)/(2N_e)$ and $r=4p/N_e$ in terms of the asymptotic flattened large-field potential $V\sim \phi^p$ can be significantly changed. Both red- and blue-shifts of $n_s$ and both enhancement and suppression of the resulting $r$ are attainable. Section~\ref{sec:strings} contains a short discussion of the structure of the underlying string theory constructions of axion monodromy inflation, and the resulting form of the 4D effective 2-field modulus-axion Lagrangians studied here. We conclude in section~\ref{sec:concl}.

\section{Effective field theories}\label{sec:EFTs}
Our goal is to understand the effect on the inflationary observables of a kinetic coupling between a heavy and a light scalar fields of the form
\be
\mathcal{L}/\sqrt{|g|}=\frac{1}{2}(\partial \phi_H)^2+\frac{f^2(\phi_H/\Lambda)}{2}(\partial \phi_L)^2-V(\phi_H,\phi_L)
\label{eq:model}
\ee
i.e. a ''gelaton'' type coupling \cite{Tolley:2009fg}. Some particular forms of this coupling can be found in the literature, in particular in \cite{Achucarro:2015rfa} (see also \cite{Achucarro:2010da}). Furthermore this sort of structure arises frequently in SUGRA or string constructions of cosmic inflation, as we will review in Sec. \ref{sec:strings}.

The idea is to analyse this action with single inflation in mind, that is in the regime $m_H\gg H$. One can of course determine the observational signatures of these models by performing a full two field analysis, following the methods developed in \cite{Sasaki:1995aw,Gordon:2000hv, GrootNibbelink:2001qt}. We'll see that this system can equivalently be analysed via two different EFTs that yield compatible results for the inflationary observables: one can either derive an EFT for the background evolution, in which case one obtains a low energy action of the $P(X)$ form as shown in \cite{Tolley:2009fg} (see also \cite{Burgess:2012dz,Gong:2014rna}) that ultimately allows for a reduction of the scalar speed of sound. Alternatively one can derive and EFT for the adiabatic curvature perturbations, by noting that in this regime the isocurvature perturbations are very massive and therefore decouple, finding once again a sub-luminal speed of sound for the scalar perturbations \cite{Achucarro:2010da}. We will show below that both EFTs yield the same results if one solves the theory as perturbative series in the small parameter $\dot{\phi}_L$.\footnote{We thing of these as top down EFTs, where one knows the UV action and integrates out heavy degrees of freedom, unlike the EFTs of \cite{Cheung:2007st,Weinberg:2008hq} which we'd call bottom up since the starting point is to write down all operators allowed by the symmetries, while remaining agnostic about the UV structures that generate such terms.}

The equations of motion that follow from Eq. \eqref{eq:model} are:
\be
\ddot{\phi}_H+3 H \dot{\phi}_H= f(\phi_H) \partial_{\phi_H}f(\phi_H) \dot{\phi_L}^2-\partial_{\phi_H} V\ ,
\ee

\be
\ddot{\phi}_L+3 H \dot{\phi}_L+ 2\frac{ \partial_{\phi_H}f(\phi_H)}{ f(\phi_H)} \dot{\phi_L}\dot{\phi_H}=-\frac{\partial_{\phi_L}V}{f^2(\phi_H)}\ ,
\ee

\be
H^2=\frac{1}{3}\left(\frac{\dot{\phi_H}^2}{2}+f^2(\phi_H)\frac{\dot{\phi_L}^2}{2}+V\right)\ ,
\ee

\be
\dot{H}=-\left(\frac{\dot{\phi_H}^2}{2}+f^2(\phi_H)\frac{\dot{\phi_L}^2}{2}\right)\ .
\ee

We assume that $m_H\gg H$ such that the heavy field can be integrated out and are interested in slow-roll solutions to this system ($\ddot{\phi}_L\approx 0$ and $\ddot{\phi}_H\approx 0$) with negligible velocity for the heavy field $\dot{\phi}_H\approx 0$. In this regime, the equation of motion for the heavy scalar reads
\be
f(\phi_H)f'(\phi_H)\dot{\phi_L}^2\approx \partial_{\phi_H}V(\phi_H)\ .
\label{eq:eomphi1}
\ee
 Solving this equation will allow us to integrate out the heavy field. Let us define
\be
\phi_H\equiv\phi_0(1+\delta)\ ,
\label{eq:phiH}
\ee
where $\phi_0$ denotes the post-inflationary minimum for $\phi_H$ and we assume that  throughout the dynamics $\delta\ll1$. This assumption is justified since the mass of $\phi_H$ is taken to be above $H$. Expanding
\be
\delta=\sum_{n\ge0} a_{2n}\dot{\phi_L}^{2n} 
\label{eq:delta}
\ee
one can solve Eq. \eqref{eq:eomphi1} order-by-order in the velocity of the light field.
Expanding both sides in powers of  $\dot{\phi}_L$ yields
\beqa
f_\alpha f'_\alpha \dot{\phi}_L^2+\frac{\phi_0 a_2}{\Lambda}\left(f'^2_\alpha+f_\alpha f''_\alpha\right){\dot{\phi}_L}^4+\mathcal{O}\left({\dot{\phi}_L}^6\right)&=&V'_\alpha+\phi_0 a_2 V''_\alpha\ \dot{\phi}_L^2\nonumber\\
&&  +\phi_0\left(a_4 V''_\alpha+\frac{1}{2}\phi_0 a_2^2 V'''_\alpha\right)\ \dot{\phi}_L^4\\
&&+\mathcal{O}\left({\dot{\phi}_L}^6\right)\quad,\nonumber
\label{eq:aux}
\eeqa
where we have used the notation $f_\alpha^{(n)}\equiv f^{(n)}(\phi_0(1+a_0))$, $V_\alpha^{(n)}\equiv V^{(n)}(\phi_0(1+a_0))$ . From Eq. \eqref{eq:aux} one finds the expressions for the $a_{2n}$
\be
\begin{split}
&V'\left(\phi_0(1+a_0),\phi_L\right)=0 \ , \\
&a_2=\frac{f_\alpha f'_\alpha  }{\phi_0 V''_\alpha \Lambda} \ ,\\
&a_4=f_\alpha f_\alpha'\frac{2 f_\alpha'^2 V_\alpha''+2 f_\alpha f_\alpha'' V_\alpha''-\Lambda f_\alpha f_\alpha' V_\alpha''' }{2 \Lambda^3 \phi_0 V_\alpha''^3}\ .
\label{eq:cnSol}
\end{split}
\ee
The first equation of \eqref{eq:cnSol} constitutes an implicit definition of $a_0$ and implies that, to leading order, the heavy field adiabatically follows its inflaton dependent minimum. Let us stress that this procedure can be carried to arbitrarily high order in $\dot{\phi}_L$ and, as we will now show, it allows us to describe the evolution of the system in terms of two distinct but equivalent EFTs.

\subsection{Background EFT}\label{sec:bEFT}

One can use the  solution to the heavy field's equation of motion, Eqs. \eqref{eq:phiH}, \eqref{eq:delta} and \eqref{eq:cnSol}, to integrate it out and find the effective field theory for the light scalar degree of freedom at the level of the background.

In the regime where one can ignore the contribution of the heavy field's kinetic term, the effective action takes the form
\be
\mathcal{L}/\sqrt{|g|}= f_\alpha^2 \frac{\dot{\phi_L}^2}{2}-V(\phi_L)+ \frac{f_\alpha^2 {f'_\alpha}^2}{2 \Lambda^2 V''_\alpha}\dot{\phi_L}^4+\frac{f_\alpha^2 {f'_\alpha}^2({3 f'_\alpha}^2 V''_\alpha+3 f_\alpha f''_\alpha V''_\alpha-\Lambda f_\alpha f'_\alpha V'''_\alpha)}{2  \Lambda^2 V''^3_\alpha}\dot{\phi_L}^6+ \mathcal{O}(\dot{\phi}_L^8)\ ,
\label{eq:PofX1}
\ee
In Eq. \eqref{eq:PofX1} it is understood that $V(\phi_L)=V(\phi_H|_{V'_\alpha=0},\phi_L)$. One therefore sees that at low energies, the inflaton $\phi_L$ has an action that is of the K-flation \cite{ArmendarizPicon:1999rj,Garriga:1999vw} or $P(X,\phi_L)$ form as noted, in the absence of mixing in $V$,  in \cite{Tolley:2009fg}. Interactions between the heavy and light scalars generically give rise not only to HD terms but also corrections to the scalar potential and kinetic terms of the light field.

The inflationary perturbations that follow from such type of action have been studied in \cite{Garriga:1999vw}. The main feature that arises from the HD terms is the reduction of the speed of sound for the scalar perturbations. Defining $X\equiv \frac{{\dot{\phi_L}}^2}{2}$ it is given by
\be
c_s^{-2}=1+2 X P_{XX}/P_{X}
\ee
and the first terms in this expansion take the form
\be
c_s^{-2}=1+ 4 \frac{ {f'_\alpha}^2}{ \Lambda^2 V''_\alpha } \dot{\phi_L}^2+4 \frac{f_\alpha'^2\left(f_\alpha'^2 V''_\alpha + 3 f_\alpha f_\alpha'' V''_\alpha-\Lambda f_\alpha f'_\alpha V'''_\alpha\right)}{V''^3_\alpha \Lambda^4}\dot{\phi_L}^4+\mathcal{O}(\dot{\phi}_L^6)\ .
\label{eq:csEFTb}
\ee

The scalar spectrum has an amplitude given by \cite{Garriga:1999vw}
\be
\mathcal{A}_s=\frac{H^2}{8 \pi^2 \epsilon c_s}\Big|_{c_s k = aH}\ ,
\label{eq:As}
\ee
with a tilt 
\be
n_s=1-2 \epsilon-\eta-s\ ,
\label{eq:ns}
\ee
where
\be
s=\frac{\dot{c_s}}{H c_s}\ .
\ee
The tensor to scalar ratio, by virtue of Eq. \eqref{eq:As} and of the fact that tensor perturbations are unaffected by the HD corrections to the scalar sector, is given by
\be
r=16 \epsilon c_s\ .
\label{eq:r}
\ee

\subsection{Perturbation EFT}\label{sec:pEFT}

An alternative approach to the one presented above is to define an EFT for the scalar perturbations, whenever there is a hierarchical mass spectrum \cite{Cremonini:2010ua,Achucarro:2010da,Achucarro:2010jv}. In what follows we sketch how this can be done, leaving out some of the details for the sake of brevity and directing the reader to \cite{Achucarro:2010da}.

The scalar perturbations are described in terms of the canonically normalised gauge-invariant Mukhanov-Sasaki variables \cite{Sasaki:1986hm,Mukhanov:1988jd}
\be
 v^a\equiv a Q^a  = a\left( \delta \phi^a + \frac{\dot{\phi}^a}{H}\psi\right).
 \ee
In two field models these can be projected in the direction parallel and perpendicular to the background trajectory
\be
v^T  =  a T_a Q^a \qquad \text{and}\qquad v^N =  a N_a Q^a\ ,
\ee
where 
\be
T = \frac{1}{|\dot{\phi}_0|} (\dot{\phi}_0^1,\quad \dot{\phi}_0^2)\quad , \quad N = \frac{1}{\sqrt{G}|\dot{\phi}_0|}(-G_{22}\dot{\phi}_0^2-G_{12}\dot{\phi}_0^1,\quad G_{11}\dot{\phi}_0^1+G_{12}\dot{\phi}_0^2)\ .
\label{eq: basis}
\ee
are the tangent and orthogonal unit vectors w.r.t. the field space metric $G$ and $|\dot{\phi_0}|\equiv\sqrt{G_{ab}\dot{\phi_0^a}\dot{\phi_0^b}}$.

The quadratic action for the scalar perturbations is that of a coupled system of two harmonic oscillators. The equations of motion for the corresponding Mukhanov-Sasaki variables written in conformal time ($d\tau=1/a(t) dt$) and  in Fourier space take the form \cite{Achucarro:2010da}:
\begin{align}
\frac{d^2 v^T_{\alpha}}{d\tau^2} + 2 a H \eta_{\perp} \frac{dv_{\alpha}^N}{d\tau}-a^2 H^2 \eta^2_{\perp}v_{\alpha}^T+\frac{d(aH\eta_{\perp})}{d\tau}v_{\alpha}^N+\Omega_{TN}v_{\alpha}^N+(\Omega_{TT}+k^2)v_{\alpha}^T &= 0, \label{eq:MStau1}\\
\frac{d^2 v^N_{\alpha}}{d\tau^2} - 2 a H \eta_{\perp} \frac{dv_{\alpha}^T}{d\tau}-a^2 H^2 \eta^2_{\perp}v_{\alpha}^N-\frac{d(aH\eta_{\perp})}{d\tau}v_{\alpha}^T+\Omega_{TN}v_{\alpha}^T+(\Omega_{NN}+k^2)v_{\alpha}^N &= 0. 
\label{eq:MStau2}
\end{align}
where the Greek index $\alpha$ labels the quantum modes of the perturbations $\alpha=1,2$.

The mass matrix $\Omega$ of Eqs. \eqref{eq:MStau1} and \eqref{eq:MStau2} is the fundamental quantity in the definition of an EFT for the perturbations. It has the following elements:
\begin{align}
\Omega_{TT}\quad &= \quad -a^2H^2 (2+2\epsilon-3\eta_{\parallel}+\eta_{\parallel}\xi_{\parallel}-4\epsilon\eta_{\parallel}+2\epsilon^2-\eta^2_{\perp})\ , \\
\Omega_{NN}\quad &= \quad -a^2H^2(2-\epsilon)+a^2 V_{NN}+a^2H^2\epsilon R\ , \\
\Omega_{TN} \quad &= \quad a^2 H^2 \eta_{\perp}(3+\epsilon-2\eta_{\parallel}-\xi_{\perp})\ .
\end{align}
Before we proceed some notation must be introduced. The second slow roll parameters, measuring the tangential and normal acceleration of the background trajectory are given by

\be
\eta_\parallel=- \frac{\ddot{\phi_0}}{H \dot{\phi_0}}\qquad\text{and}\qquad\eta_\perp= \frac{V_N}{|\dot{\phi_0}| H}
\ee
respectively.  The third slow-roll parameters in turn are defined as:
\be
\xi_{\parallel} = - \frac{\dddot{\phi}_0}{H \ddot{\phi}_0}\qquad ,\qquad\xi_{\perp}  = -\frac{\dot{\eta}_{\perp}}{H \eta_{\perp}} \ .
\ee
The Ricci scalar of the scalar manifold is denoted by $R$.

Whenever there is a hierarchy
\be
|\Omega_{NN}|\gg|\Omega_{TT}|\ ,\ |\Omega_{TN}|
\ee
the two field system of Eqs. \eqref{eq:MStau1}-\eqref{eq:MStau2} can be equivalently described by an EFT for the light degree of freedom $v^T$. For large $\Omega_{NN}$ one can solve Eq. \eqref{eq:MStau2} in the limit of negligible acceleration finding
\be
v^N=\frac{2 a H \eta_\perp \frac{d v^T}{d \tau}+\frac{d (a H \eta_\perp)}{d \tau} v^T-\Omega_{NT} v^T}{\Omega_{NN}-(a H \eta_\perp)^2+k^2}.
\ee
Substituting this in Eq. \eqref{eq:MStau1} and defining the canonical variable $u$
\be
v_T=e^{-\beta/2} u,
\ee
one finds that the curvature perturbations follow the evolution equation
\be
u''+ (c_s^2 k^2 +\Omega )u=0,
\ee
propagating at the speed of sound \cite{Achucarro:2010jv,Achucarro:2010da}
\be
c_s^{2}=e^{-\beta}=\frac{k^2- a H \eta_\perp+\Omega_{NN}}{k^2+3 a H \eta_\perp+\Omega_{NN}}
\label{eq:etaPerp}
\ee
and with an effective mass
\be 
 \Omega= \Omega_{TT}- (a H \eta_\perp)^2- a H \beta' (1+\epsilon-\eta_\parallel)-\left(\frac{\beta'}{2}\right)^2-\frac{\beta''}{2}\ ,
 \ee
to leading order in $\Omega_{NN}$. The validity and accuracy of this EFT has also been considered  in e.g. \cite{Avgoustidis:2011em, Cespedes:2012hu, Achucarro:2012yr}.

Let us now apply this approach to the system of Eq. \eqref{eq:model}. Neglecting the velocity for the heavy field, $\dot{\phi_H}\approx 0$, one writes
\be
T^a=\frac{1}{f(\phi_H)}(0,1) \qquad\text{and}\qquad N^a=(-1,0),
\ee
in which case
\be
\eta_\perp= -\frac{V'(\phi_H) }{|\dot{\phi_L}| H f(\phi_H /\Lambda)}
\label{eq:etaPerp}
\ee
implying a speed of sound 
\be
c_s^{-2}=1+4 \eta_\perp^2 \left(\frac{V''(\phi_H )}{H^2}-\frac{\dot{\phi_L}^2}{H^2} f''(\phi_H) f(\phi_H)-\eta_\perp^2\right)^{-1}.
\label{eq:csMinus2P}
\ee
Note that in the above derivation we have used the fact that the field space curvature is given by $R=-2 \frac{f''(\phi_H)}{f(\phi_H)}$ in the models under consideration and neglected terms of the form $\frac{k}{aH}$, which is a good approximation on superhorizon scales.

Knowing the background evolution of the system, and in particular the series expansion of $\delta\equiv \phi_H/\phi_0-1$, one can determine the expansion of Eq. \eqref{eq:csMinus2P} in powers of $\dot{\phi_L}$. The first terms in this expansion are
\be
c_s^{-2}=1+ 4 \frac{ {f'_\alpha}^2}{\Lambda^2 V''_\alpha} \dot{\phi_L}^2+4 \frac{f_\alpha'^2  \left(f_\alpha'^2 V''_\alpha +3f_\alpha f_\alpha'' V''_\alpha-\Lambda f_\alpha f'_\alpha V'''_\alpha\right)}{\Lambda^4 V''^3_\alpha}\dot{\phi_L}^4+ \mathcal{O}(\dot{\phi}_L^6)\ .
\ee
We therefore find that the result agrees with the estimate for $c_s$ obtained from the background EFT of Sec. \ref{sec:bEFT}, Eq. \eqref{eq:csEFTb}. The agreement between the two estimates for $c_s$ will hold to order  $2n$ in $\dot{\phi_L}$ if one solves Eq.\eqref{eq:eomphi1} to order $2n+2$.

Having established that the system of Eq. \eqref{eq:model} can be analysed by means of two equivalent EFTs we will now proceed to study particular examples aiming to understand under what circumstances reductions of $c_s$ are attainable .

\section{Kinetic Mixing}\label{sec:KineticMix}

In this section we study in more detail the case when there is no mixing between $\phi_L$ and $\phi_H$ in $V$.  We explicitly analyse two examples of coupling function and their impact on the estimates of the speed of sound and inflationary observables, comparing the estimates from the two EFTs defined above to the results from the full two field numerical evolution of the system.

We assume that the potential for the heavy scalar is quadratic around its minimum at $\phi_0$, implying that the scalar potential takes the form
\be
V=\frac{1}{2} m_H^2(\phi_H-\phi_0)^2+V(\phi_L)\ .
\label{eq:VnoMix}
\ee
The absence of mixing in $V$  implies that the shift of the heavy field away from its potential minimum, $\delta$, depends only on the velocity of the light field (and not on the light field itself), since from Eq. \eqref{eq:cnSol} one finds $a_0=0$. For the potential of Eq. \eqref{eq:VnoMix} this implies that $\alpha=\phi_0$, $f^{(n)}_\alpha=f^{(n)}(\phi_0/\Lambda)\equiv f^{(n)}_0$, $V^{(n)}_\alpha=V^{(n)}(\phi_0,\phi_L)$. From Eq. \eqref{eq:PofX1} one can define the canonically normalised variable via
\be
\Phi_L\equiv f_0 \phi_L\ ,
\ee
in which case the action simply reads
\be
\mathcal{L}/\sqrt{|g|}= \frac{\dot{\Phi_L}^2}{2}-V(\Phi_L)+ \frac{{f'_0}^2/f_0^2 }{2 m^2 \Lambda^2}\dot{\Phi_L}^4+\frac{ {f'_0}^2/f_0^4({f'_0}^2+f_0 f''_0)}{2 m^2 \Lambda^2}\dot{\Phi_L}^6+...\ .
\ee
 We therefore conclude that in the absence of mixing in $V$ there are no corrections to the scalar potential or to the kinetic term, the only feature that remains from the generic analysis of the previous section are HD terms in the form of higher powers of first derivatives. This is the standard gelaton scenario of \cite{Tolley:2009fg}.

Knowing that a reduction of $c_s$ entails a reduction of the tensor-to-scalar ratio, we focus our analysis on the class of chaotic monomials and in particular of quadratic inflationary potential:
\be
V(\phi_L)=\frac{1}{2}m_L^2 \phi_L^2\ .
\ee
A phenomenological analysis of the impact of $c_s$ reductions attainable in this class of models can be found in \cite{Stein:2016jja}.

We work in a regime where $m_H = 10 H$ (at horizon exit of CMB scales), following \cite{Achucarro:2015rfa} whose results we partially reproduce in Sec. \ref{sec:quadratic}.
For concreteness we solve Eq. \eqref{eq:eomphi1} to 12th order in $\dot{\phi_L}$ which as we'll demonstrate allows the use of the EFTs  all the way down to $c_s\sim 0.6.$\footnote{Having some knowledge of the UV physics allows us to go beyond the modest reductions of $c_s$ one can reliably find when working to next-to-leading order in the derivative expansion as in e.g. \cite{Pedro:2017qcx}.}

\subsection{Monomial coupling} \label{sec:quadratic}

The first class of kinetic couplings we consider are monomials of the form
\be
f^2(\phi_H)=\left(\frac{\phi_H}{\Lambda}\right)^{2p}\ .
\ee
The speed of sound for the scalar perturbations can be written as 
\be
c_s^{-2}=1+ \sum_{n\ge1} a_{2n} \left(\frac{\dot{\phi_L} (\phi_0/\Lambda)^p}{m_H \phi_0}\right)^{2n},
\label{eq:}
\ee
where the first four numerical coefficients take the form
\beqa
a_2&=& 4 p^2 \ , \nonumber\\
 a_4&=& 4 p^3(-3+4p) \ ,\nonumber\\
  a_6&=&4 p^4(10-27p+18p^2)\ ,\nonumber\\
   a_8&=&\frac{4}{3}p^5 (-105 + 428 p - 576 p^2 + 256 p^3) \ . 
\label{eq:coeffMon}
\eeqa
Noting that in the background EFT the canonically normalised inflaton is defined by $\Phi\equiv\phi_L (\frac{\phi_0}{\Lambda})^p$ one sees that 
\be
c_s^{-2}-1\propto \frac{{\dot{\Phi}^2}}{m_H^2 \phi_0^2}\sim\epsilon \frac{H^2}{m_H^2}\frac{M_P^2}{\phi_0^2}
\ee
and so reductions of $c_s$ for fixed $m_H/H$ can only be achieved via the tuning of $\phi_0$ to small values.

In order to be more explicit and to gauge the validity of the EFT descriptions  one must choose a particular value for $p$. In what follows we consider a quadratic coupling, $p=1$, that may arise in supergravity theories if the two scalars are the modulus and the angle of a given complex scalar field. We note however that qualitatively similar results can be found for other monomial couplings.

The speed of sound for the scalar perturbations can be written as 
\be
c_s^{-2}=1+ \sum_{n=1}^6 a_{2n} \left(\frac{\dot{\phi_L}}{m_H \Lambda}\right)^{2n},
\ee
where the $a_{2n}$ coefficients for the background EFT are found to be
\be
a_2= 4 \ , \ a_4= 4 \ ,\ a_6=4 \ , \ a_8=4 \ ,\ a_{10}= 4\ ,\  a_{12}= 4
\label{eq:a_2nb}
\ee
and for the perturbation EFT are
\be
a_2= 4 \ , \ a_4= 4 \ ,\ a_6=4 \ , \ a_8=4 \ ,\ a_{10}= 4\ ,\  a_{12}= -4 \ .
\label{eq:a_2np}
\ee
For this simple case it is possible to go beyond the perturbative method described above and solve the equation of motion for the heavy scalar, Eq. \eqref{eq:eomphi1}, exactly (in the limit of negligible velocity and acceleration).  By doing so we can gauge the precision of the approach described above. The solution to Eq. \eqref{eq:eomphi1} is
\be
\phi_H= \frac{\phi_0}{1-\frac{\dot{\phi_L}^2}{m_H^2 \Lambda^2} } \ ,
\label{eq:phiHquad}
\ee
which implies the following effective action for the background
\be
\frac{\mathcal{L}\big|_{EFT b}}{\sqrt{|g|}}= \frac{1}{\left(1-\frac{\dot{\Phi}_L^2}{m_H^2 \phi_0^2}\right)}\frac{\dot{\Phi}_L^2}{2}-V(\Phi_L)
\ee
and a sound speed given by
\be
c_s^{-2}\big|_{EFTb}= \frac{m_H^2 \phi_0^2+3 \dot{\Phi}_L^2}{m_H^2 \phi_0^2- \dot{\Phi}_L^2}\ .
\label{eq:csExact}
\ee
For the perturbation EFT of Sec. \ref{sec:pEFT} one can substitute Eq. \eqref{eq:phiHquad} in Eqs. \eqref{eq:etaPerp} and \eqref{eq:csMinus2P} to find that the speed of sound given exactly by Eq. \eqref{eq:csExact} as found in \cite{Achucarro:2015rfa}. So in this simple case, where the heavy field can be integrated out exactly\footnote{Note that we are still neglecting the effects of the velocity of the heavy field and that Eqs. \eqref{eq:phiHquad} -\eqref{eq:csExact} are to be understood as a power series in $\dot{\Phi}_L$.}, one can show that the two EFTs are not just approximately equivalent but are equivalent to all orders in the $\dot{\Phi}_L$  expansion. Furthermore, noting that Eq. \eqref{eq:csExact} can be written in the form 
\be
c_s^{-2}=1+ 4 \sum_{n=1}^\infty  \left(\frac{\dot{\phi_L}}{m_H \Lambda}\right)^{2n},
\ee
one can extrapolate, by comparison with Eqs. \eqref{eq:a_2nb} and \eqref{eq:a_2np}, that solving \eqref{eq:eomphi1} to order $2n+2$ results in $c_s^{-2}\big|_{EFTb}$ that is accurate to order $\dot{\phi_L}^{2n+2}$ and $c_s^{-2}\big|_{EFTp}$ that is accurate to order  $\dot{\phi_L}^{2n}$.

 Since the scalar potential can be written as
\be
V=\frac{m_L^2}{2}\phi_L^2=\frac{(m_L \Lambda/\phi_0)^2}{2}{\Phi_L}^2\ ,
\ee
we see that tuning $\phi_0$ small increases the mass of the canonically normalised inflaton, bringing it closer to $H$ and to $m_H$.

\begin{figure}[t!]
	\centering
	\begin{minipage}[b]{0.4\linewidth}
	\centering
	\includegraphics[width=\textwidth]{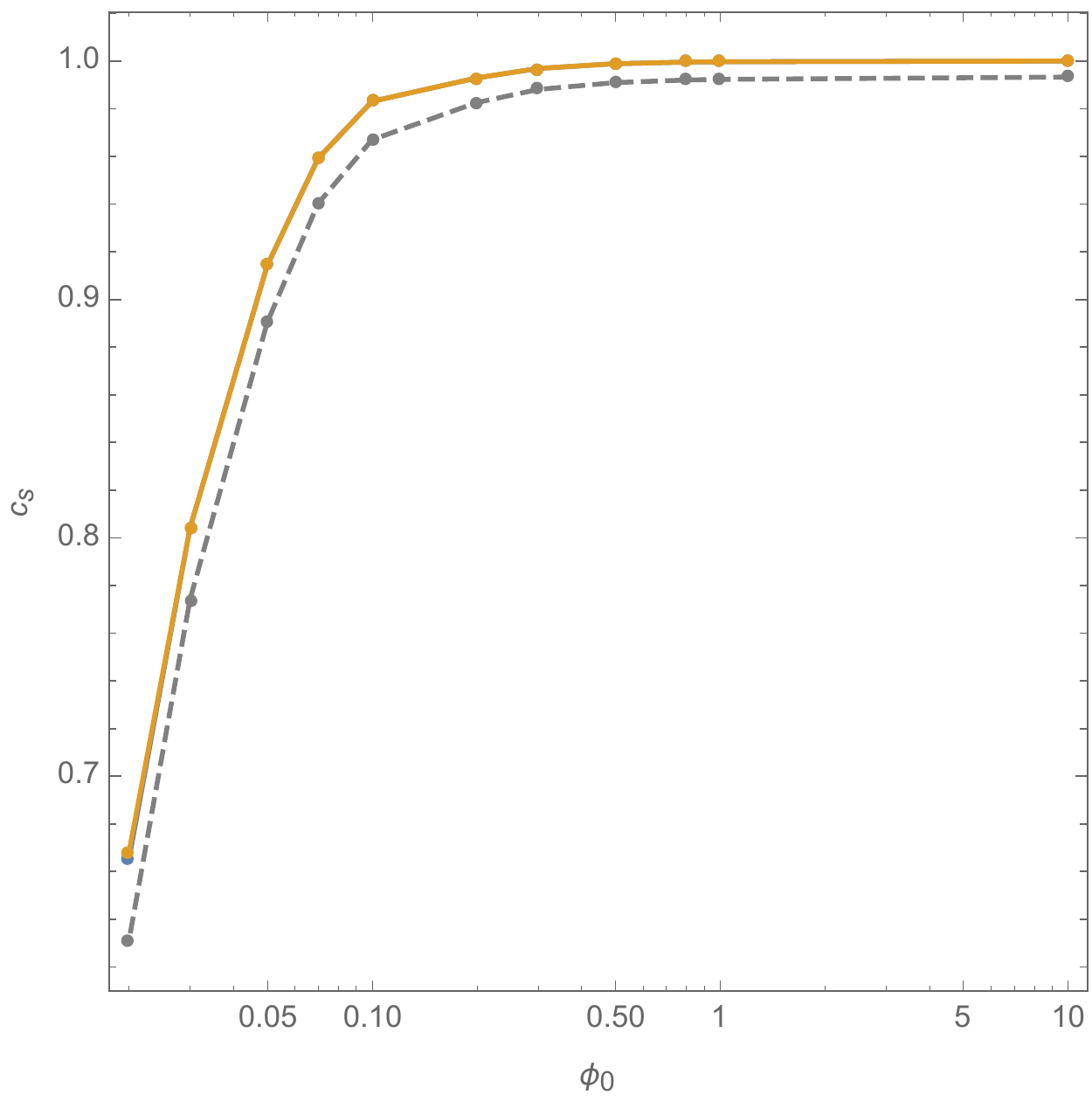}
    \end{minipage}
	\hspace{1cm}
	\begin{minipage}[b]{0.41\linewidth}
	\centering
	\includegraphics[width=\textwidth]{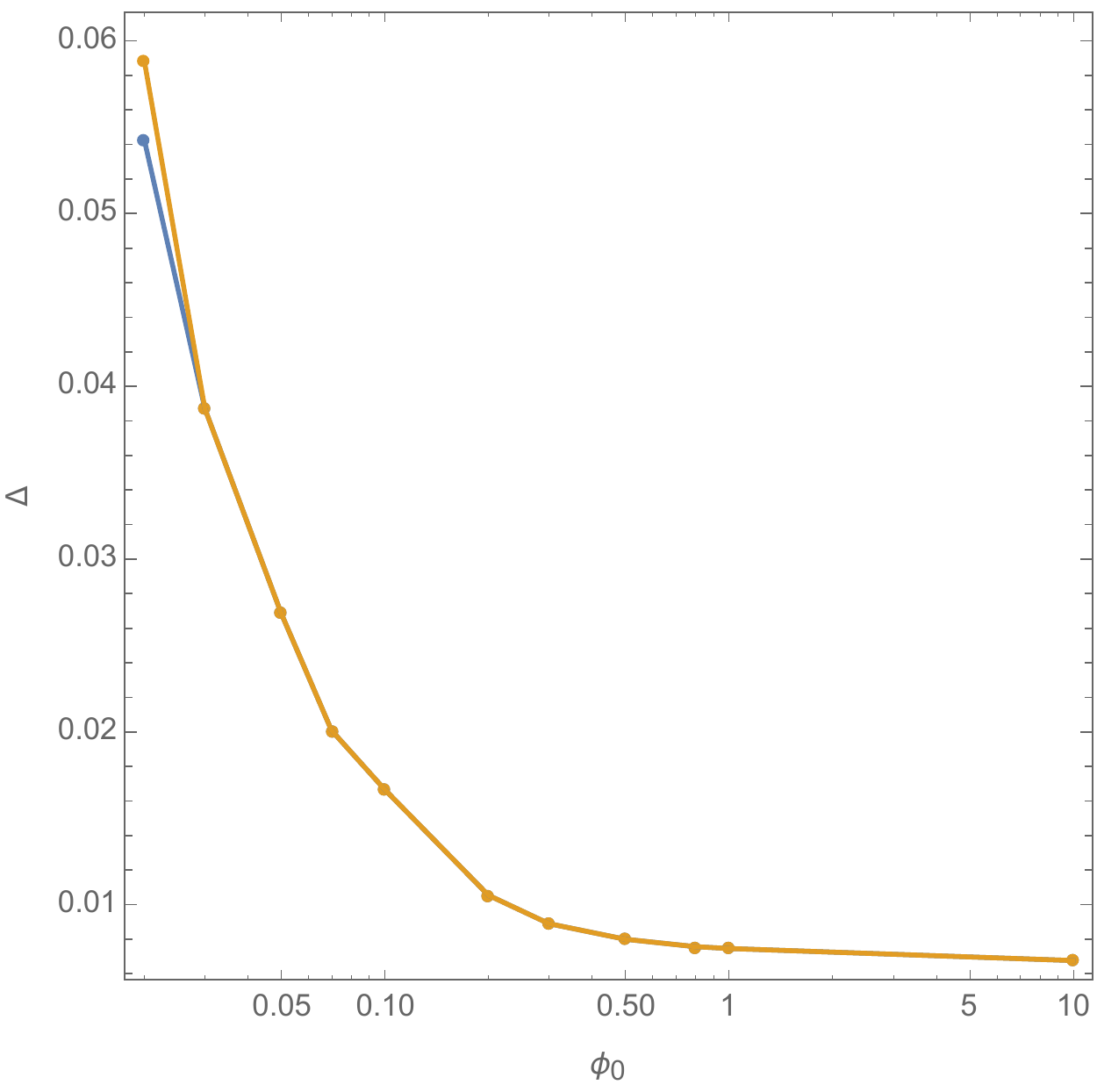}
	\end{minipage}
	\hspace{0.05cm}
	\caption{{\it Left}:  Speed of sound for a quadratic coupling. Grey dashed line represents the exact result, yellow depicts the background EFT estimate and blue the perturbation EFT. {\it Right}: Error in the EFT estimates of the speed of sound relative to the numerical solution of the two field system. }
	\label{fig:quadCoupling}
\end{figure}

In Fig. \ref{fig:quadCoupling} we plot $c_s$ as determined by the two EFTs and by the solution to the 2 field perturbation system, Eqs. \eqref{eq:MStau1}-\eqref{eq:MStau2}, and the error of the two EFTs w.r.t. the full two field result, defined as
\be
\Delta=\frac{c_s|_{EFT}-c_s}{c_s}\ .
\ee
We see that the two EFTs are in excellent agreement with each other and that they capture the full 2 field (background plus perturbation) evolution represented by the dashed line in Fig. \ref{fig:quadCoupling}. The errors of the EFT estimates increase as $c_s$ decreases, though its absolute values are still at the percent level.

The spectral index remains essentially unchanged while the tensor-to-scalar ratio decreases due to the reduction in $c_s$, in accordance with   \eqref{eq:r}. These results are in line with those of \cite{Achucarro:2015rfa}.

\subsection{Exponential coupling}\label{sec:exp1}
\begin{figure}[t!]
	\centering
	\begin{minipage}[b]{0.4\linewidth}
	\centering
	\includegraphics[width=\textwidth]{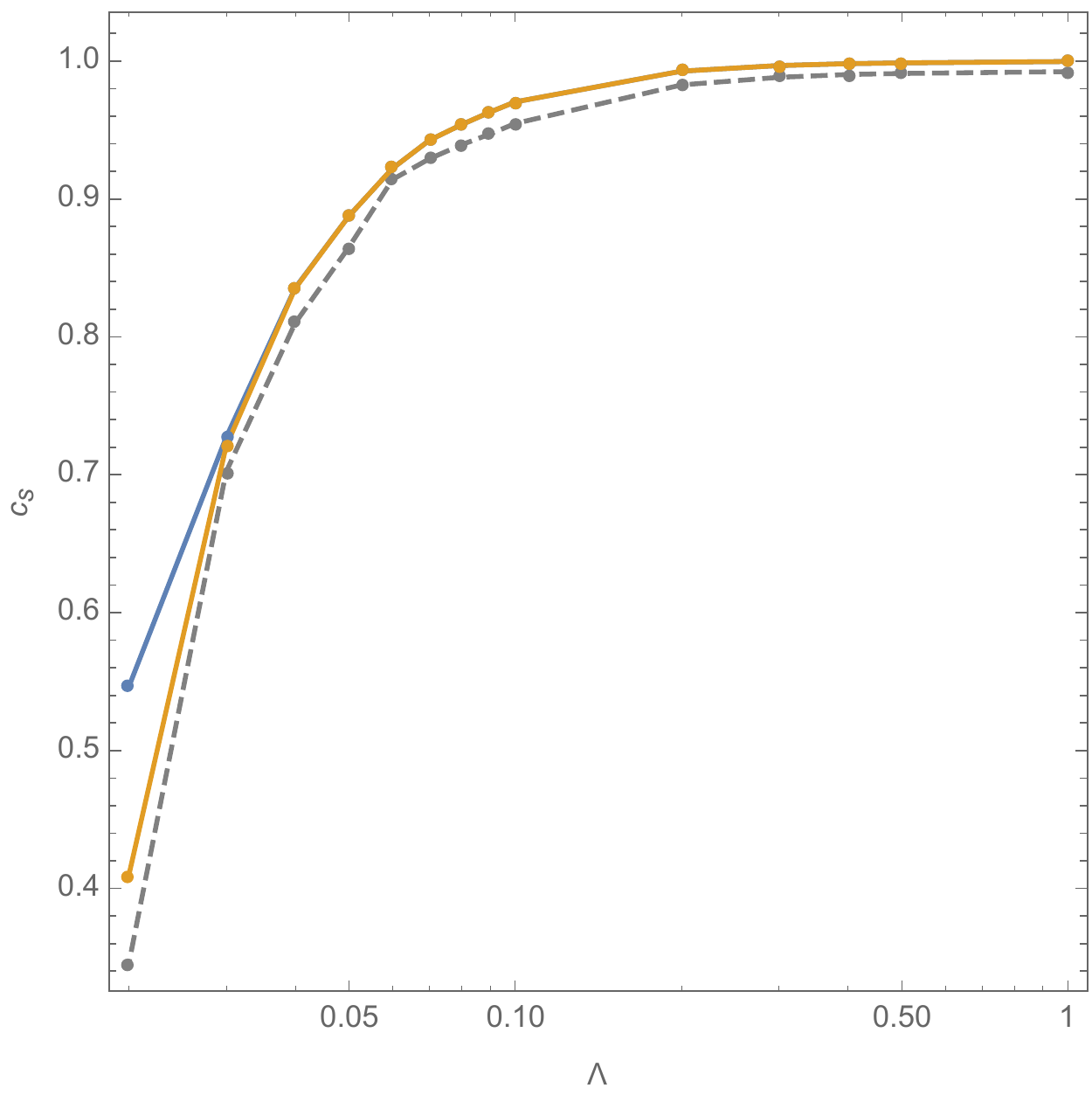}
    \end{minipage}
	\hspace{1cm}
	\begin{minipage}[b]{0.41\linewidth}
	\centering
	\includegraphics[width=\textwidth]{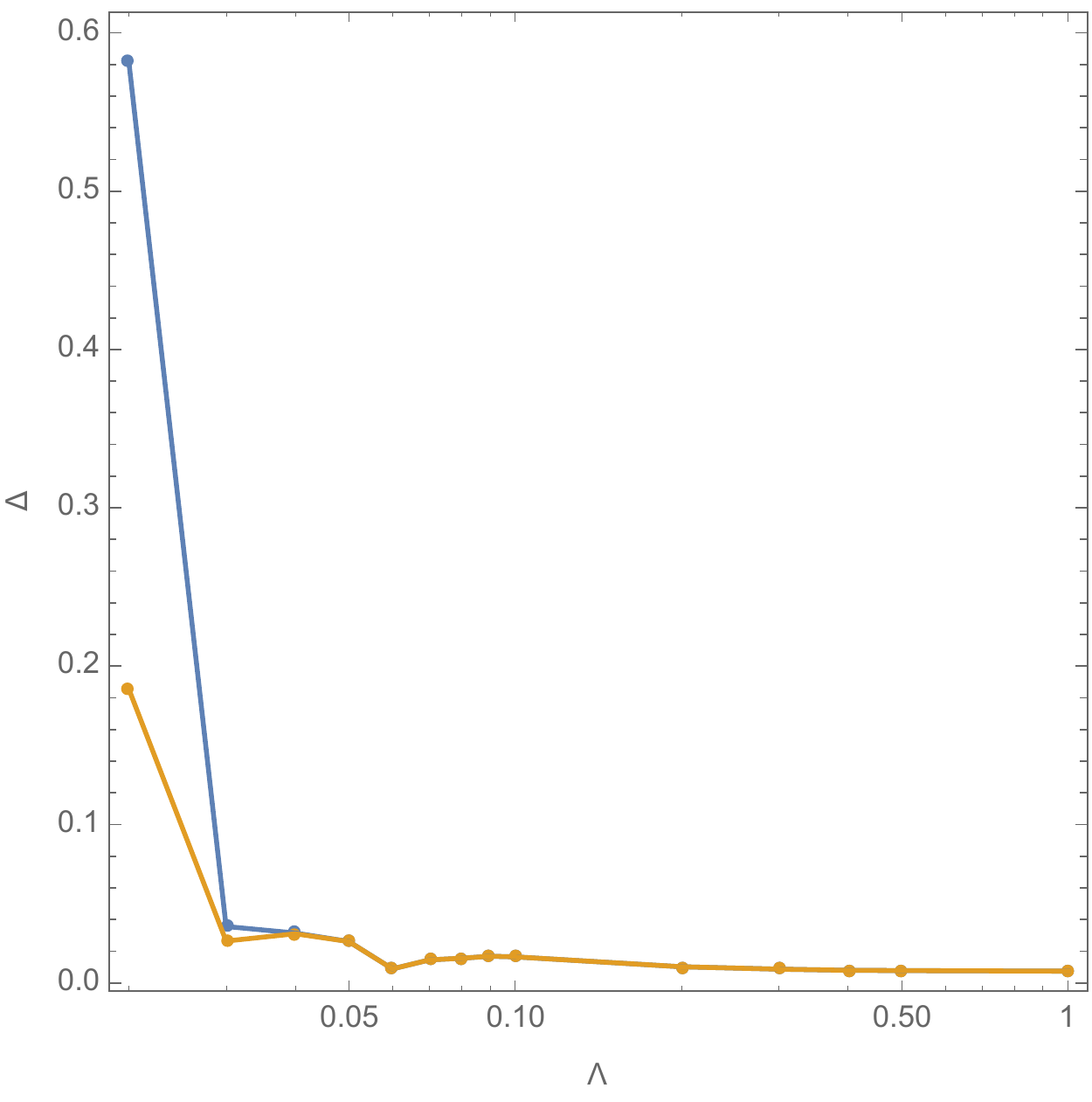}
	\end{minipage}
	\hspace{0.05cm}
\caption{{\it Left}:  Speed of sound for a exponential coupling. Grey dashed line represents the exact result, yellow depicts the background EFT estimate and blue the perturbation EFT. {\it Right}: Error in the EFT estimates of the speed of sound relative to the numerical solution of the two field system. }
	\label{fig:expCoupling}
\end{figure}

Let us now consider an exponential coupling function
\be
f^2(\phi_H)=\exp\left( 2\frac{\phi_H}{\Lambda}\right),
\ee
which, as we will discuss in Sec. \ref{sec:strings}, often appears in UV embeddings of inflation.
This yields a speed of sound of the form
\be
c_s^{-2}=1+ \sum_{n=1}^6 a_{2n} \left(\frac{\dot{\phi_L} }{m_H \Lambda e^{- \phi_0/\Lambda}}\right)^{2n}\ ,
\label{eq:csExp}
\ee
where for the background EFT one finds
\be
a_2=  4\ , \ a_4= 16 \ ,\ a_6=72 \ , \ a_8=1024 \ ,\ a_{10}=5000 \ ,\  a_{12}= 41472 \ ,
\ee
while for the perturbation EFT the expansion is 
\be
a_2=  4\ , \ a_4= 16 \ ,\ a_6=72 \ , \ a_8=1024 \ ,\ a_{10}=5000 \ ,\  a_{12}= 27648 \ .
\ee
Writing Eq. \eqref{eq:csExp} in terms of the canonical variable one obtains
\be
c_s^{-2}-1\propto \frac{{\dot{\Phi}_L^2}}{m_H^2 \Lambda^2}\sim \epsilon \frac{H^2}{m_H^2} \frac{M_p^2}{\Lambda^2}\ ,
\ee
implying that reductions of $c_s$ for fixed $H/m_H$ may be obtained via tuning of $\Lambda$. In Fig. \ref{fig:expCoupling} we compare the various estimates for $c_s$ as functions of the scale $\Lambda$. The qualitative picture is similar to what was found in the quadratic coupling case analysed above: significant reductions of $c_s$ are attainable in the small $\Lambda$ limit, with  little impact on the tilt of the scalar power spectrum.

\section{Kinetic and potential mixing}\label{sec:PotKinMix}

Let us now consider the generic case when there is mixing in the scalar potential as well as in the kinetic terms. In this case the background EFT features non-canonical kinetic terms for the inflaton field, since $f_\alpha=f_\alpha(\phi_L)$, a direct consequence of the fact that the vev of the heavy field depends both on the light field and its derivative. 

Let us consider the effective action for the light field, Eq. \eqref{eq:PofX1}, and write it in terms of the canonical variable $\Phi_L$, defined as 
\be
\Phi_L=\int f_\alpha(\phi_L) d\phi_L\ .
\ee
One may then write the effective Lagrangian as
\be
\begin{split}
\mathcal{L}/\sqrt{|g|}=& \frac{(\partial \Phi_L)^2}{2}-V_{eff}(\phi_L(\Phi_L))+\\
&+\frac{1}{2}(\partial \Phi_L)^4 \frac{(f'_\alpha (\Phi_L)/f_\alpha(\Phi_L))^2}{\Lambda^2(m^2+ g \phi_L^2(\Phi_L))}+\frac{1}{2}(\partial \Phi_L)^6 \frac{(f'_\alpha (\Phi_L)^4 + f_\alpha(\Phi_L) f'^2_\alpha(\Phi_L) f''_\alpha(\Phi_L))}{\Lambda^4(m^2+ g \phi_L^2(\Phi_L))^4 f_\alpha(\Phi_L)^4}+...
\label{eq:LeffMix}
\end{split}
\ee
where the effective potential can be found from Eq.  \eqref{eq:cnSol} and takes the form
\be
V_{eff}(\phi_L(\Phi))= V(\phi_L(\Phi))+\frac{g m_H^2 \phi_0^2}{2}\frac{\phi_L^2(\Phi)}{m_H^2+g \phi_L^2(\Phi)}\ ,
\label{eq:Veff}
\ee 
if for concreteness one assumes that in the UV
\be
V=\frac{1}{2} m_H^2(\phi_H-\phi_0)^2+\frac{g}{2} \phi_H^2 \phi_L^2+V(\phi_L)\ .
\ee
Note that in the case $V(\phi_L)=0$ and in the absence of kinetic mixing, Eq. \eqref{eq:Veff} reduces to the prototypical flattened potential of \cite{Dong:2010in}, that for large values of $\phi_L$ asymptotes to a constant $V\sim m_H^2 \phi_0^2 / 2 $. The crucial thing to note is that Eq. \eqref{eq:LeffMix} features not only HD terms of the type found in the previous section, but also corrections to the kinetic term and scalar potential, Eq. \eqref{eq:Veff}. 

We can now study the same examples as in Sec. \ref{sec:KineticMix} with the added mixing in the potential and demonstrate that this modification leads to a very different behaviour of the system in what concerns inflationary dynamics and observables. In order to make contact with the results of Sec. \ref{sec:KineticMix} we will assume
\be
V(\phi_L)=\frac{1}{2}m_L^2 \phi_L^2 \ ,
\ee
however we will also study the possibility that $V(\phi_L)=0$, in which case the inflationary potential is generated by the interaction between the heavy field and the inflaton, as is often the case in UV constructions of inflation.

\subsection{Monomial kinetic coupling}

We now revisit the monomial kinetic couplings
\be
f(\phi_H)=\left(\frac{\phi_H}{\Lambda}\right)^{2p}
\ee
in the presence of an interaction between the fields in the scalar potential. One can show that the speed of sound takes the form
\be
c_s^{-2}=1+ \sum_{n\ge1} a_{2n} \left(\frac{\dot{\phi_L} }{M_H \phi_0}\left(\frac{\phi_0}{\Lambda}\right)^p\right)^{2n},
\label{eq:}
\ee
where one defined the total mass of the heavy field as $M_H^2= m_H^2+g \phi_L^2$ and the numerical coefficients, $a_{2n}$, are those of Eq. \eqref{eq:coeffMon}. This can be rewritten as 
\be
c_s^{-2}-1\approx \sum_{n\ge1} \left[ \epsilon \left(\frac{H}{M_H}\right)^2 \left(\frac{M_p}{\phi_0}\right)^2 \left(1+\frac{g \phi_L^2}{m_H^2}\right)^{2p} \right]^n
\ee
where one sees that reductions of $c_s$ may arise in the small $\phi_0/M_P$ limit. For $p>0$ the suppression of the speed of sound can be further enhanced by pushing the system into the $\frac{g \phi_L^2}{m_H^2} \ge \mathcal{O}(1)$ regime.

Let us be more concrete and analyse in detail the case $f(\phi_H)=\phi_H/\Lambda$, which has the virtue of being simple enough to treat exactly while keeping all the features of higher order monomials. For this form of the kinetic coupling one can not only perform the integration relating the canonical and non-canonical variables but also analytically invert that relation to find that the canonical variable is 
\be
\Phi= \frac{m_H \phi_0}{\sqrt{g} \Lambda} \tan^{-1}\left(\frac{\sqrt{g} \phi_L}{m_H}\right)\ .
\ee
The effective action is a function of two mass scales besides the inflaton mass $m_L$, namely 
\be
\mu^4\equiv \frac{1}{2} m_H^2 \phi_0^2 \qquad\text{and}\qquad f_a\equiv \frac{m_H \phi_0}{\sqrt{g} \Lambda}
\label{eq:mufa}
\ee
and takes the form
\be
\mathcal{L}_{eff}/\sqrt{|g|}= \frac{1}{2}\dot{\Phi}^2-V_{eff}(\Phi)+\sum_{n>1} \frac{\dot{\Phi}^{2n}}{2^n \mu^{2n-4} \cos^{2n-2} (\Phi/f_a) }\ , 
\ee
where the effective potential is given by
\be
V_{eff}(\Phi)= \frac{m_L^2 \ \mu^4}{g \phi_0^2} \tan^2\left(\frac{\Phi}{f_a}\right)+\mu^4\left[1-\cos^2\left(\frac{\Phi}{f_a} \right)\right ].
\label{eq:VeffMon}
\ee

From Eq. \eqref{eq:} one can show that the speed of sound can be written as 
\be
c_s^{-2}=1+ \sum_{n\ge 1} a_{2n} \left(\frac{m_H^2+g \phi_L^2(\Phi)}{m_H^4 \phi_0^2}\right)^n \dot{\Phi}^{2n}\ ,
\label{eq:csQuadMix}
\ee
where from the perturbative analysis of the background EFT one finds
\be
a_2=  4\ , \ a_4= 4 \ ,\ a_6=4 \ , \ a_8=4 \ ,\ a_{10}=4 \ ,\  a_{12}= -80 \ ,
\ee
whereas the EFT for the perturbations yields
\be
a_2=  4\ , \ a_4= 4 \ ,\ a_6=4 \ , \ a_8=4 \ ,\ a_{10}=4 \ ,\  a_{12}= 4 \ ,
\ee
if one solves Eq. \ref{eq:eomphi1} to 12th order in $\dot{\Phi}$.
Note that these results are compatible with those of Sec. \ref{sec:quadratic}, reducing to Eqs. \eqref{eq:a_2nb} and \eqref{eq:a_2np} in the $g\rightarrow0$ limit.

The eom for the heavy field, in the limit where one neglects its velocity and acceleration, reduces to a linear equation for $\phi_H$ one therefore can show that its solution is 
\be
\phi_H=\frac{\phi_0}{1+\frac{g \phi_L^2}{m_H^2}-\frac{\dot{\phi}_L^2}{\Lambda^2 m_H^2}} \ .
\ee
This relation is exact to all orders in $\dot{\phi}_{L}$ and it allows us to write the effective Lagrangian as
\be
\mathcal{L}_{eff}/\sqrt{|g|}=\frac{m_H^2 \phi_0^2}{2}\frac{\dot{\phi}_L^2-g \Lambda^2 \phi_L^2}{\Lambda^2(m_H^2+g \phi_L^2) -\dot{\phi}_L^2}-V(\phi_L)
\ee
from which the speed of sound is 
\be
\begin{split}
c_s^{-2}&= \frac{\Lambda^2(m_H^2 +g \phi_L^2)+3 \dot{\phi_L^2}}{\Lambda^2(m_H^2 +g \phi_L^2)- \dot{\phi_L}^2}\\
&=1+ 4 \sum_{n=1}^\infty  \left(\frac{\dot{\phi_L}^2}{\Lambda^2 (m_H^2 +g \phi_L^2) }\right)^{n}\ .
\end{split}
\ee
Writing $c_s$ in terms of the canonical variable one finds Eq. \eqref{eq:csQuadMix} with  $a_{2n}=4,\  \forall \  n$.
This again confirms that if one expands the action to order $2n+2$ in $\dot{\phi}_L$ one can expect the results to be accurate to order $2n$.

\paragraph{Regime 1: $m_L^2\gg g \phi_0^2$\\}
In this regime the inflationary potential is dominated by the self-interactions of the inflaton field, with the heavy field making its presence felt through the kinetic interaction/ canonical normalisation of the inflaton.  Given that we are interested in working in the regime $\phi_0/M_P\ll1$ we conclude that in this regime $\frac{m_H^2}{g \phi_L^2}\gg1$ and that therefore $M_H=m_H$.
The non-canonical kinetic term for $\phi_L$ leads to a steepening of the potential for the canonical variable $\Phi$, which  can schematically be written as 
\be
V\propto \Phi^2\left(1+\sum_{n\ge1} \nu_{2n} \left(\frac{\Phi}{f_a}\right)^{2n}\right)
\ee
with  coefficients $\nu_{2n}>0\quad  \forall \ n$. Such steepening will occur whenever $f(\phi_H(\phi_L))$ scales as a negative power of $\phi_L$ in the relevant range for inflationary dynamics, a fact observed already in \cite{Dong:2010in}. This effect will turn out to be crucial for the determination of the inflationary observables and to the extent to which one can decrease the speed of sound while being in line with observational constraints on the remaining observables.

\begin{figure}[t!]
	\centering
	\begin{minipage}[b]{0.3\linewidth}
	\centering
	\includegraphics[width=\textwidth]{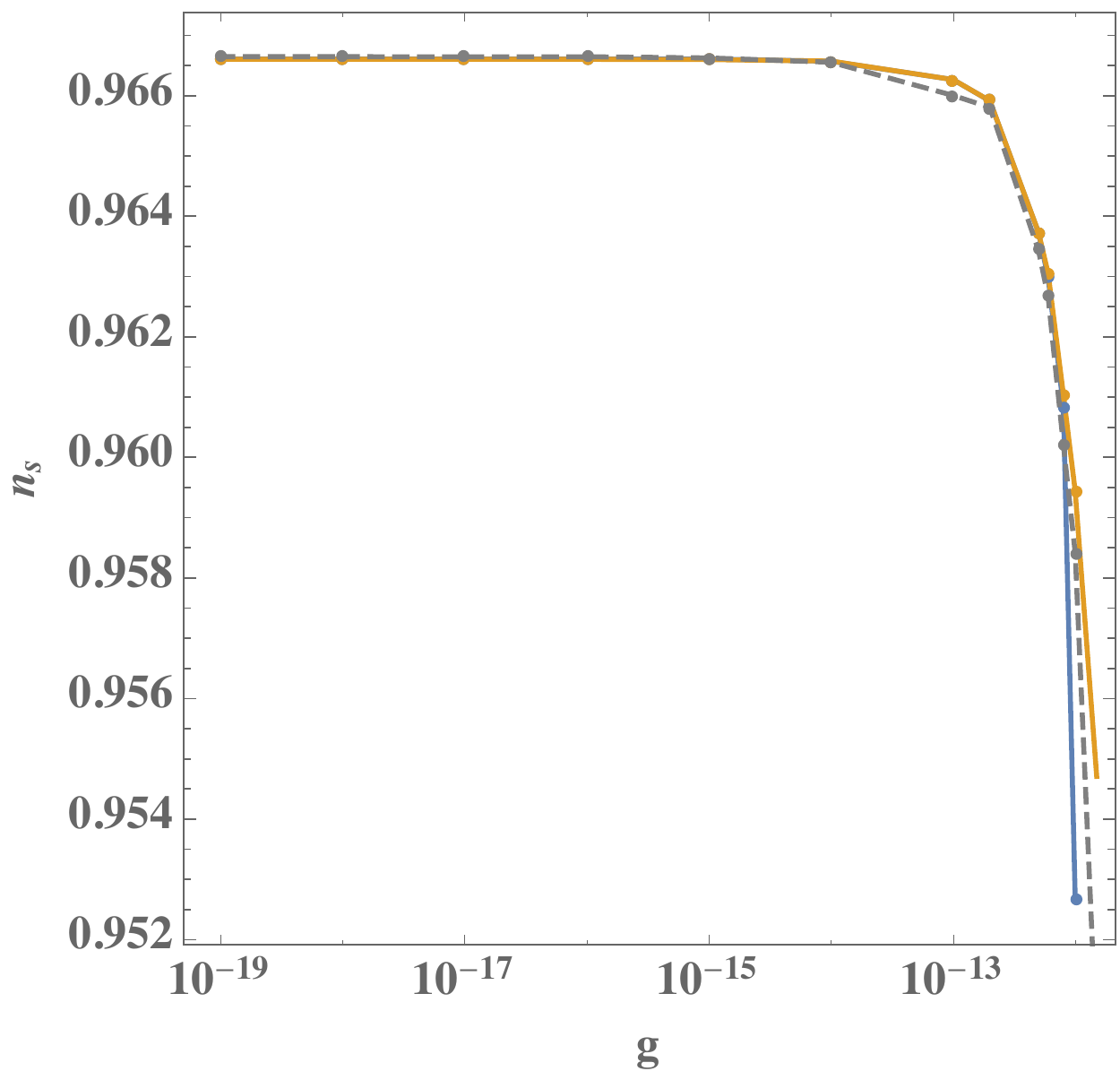}
    \end{minipage}
	\hspace{0.5cm}
	\begin{minipage}[b]{0.3\linewidth}
	\centering
	\includegraphics[width=\textwidth]{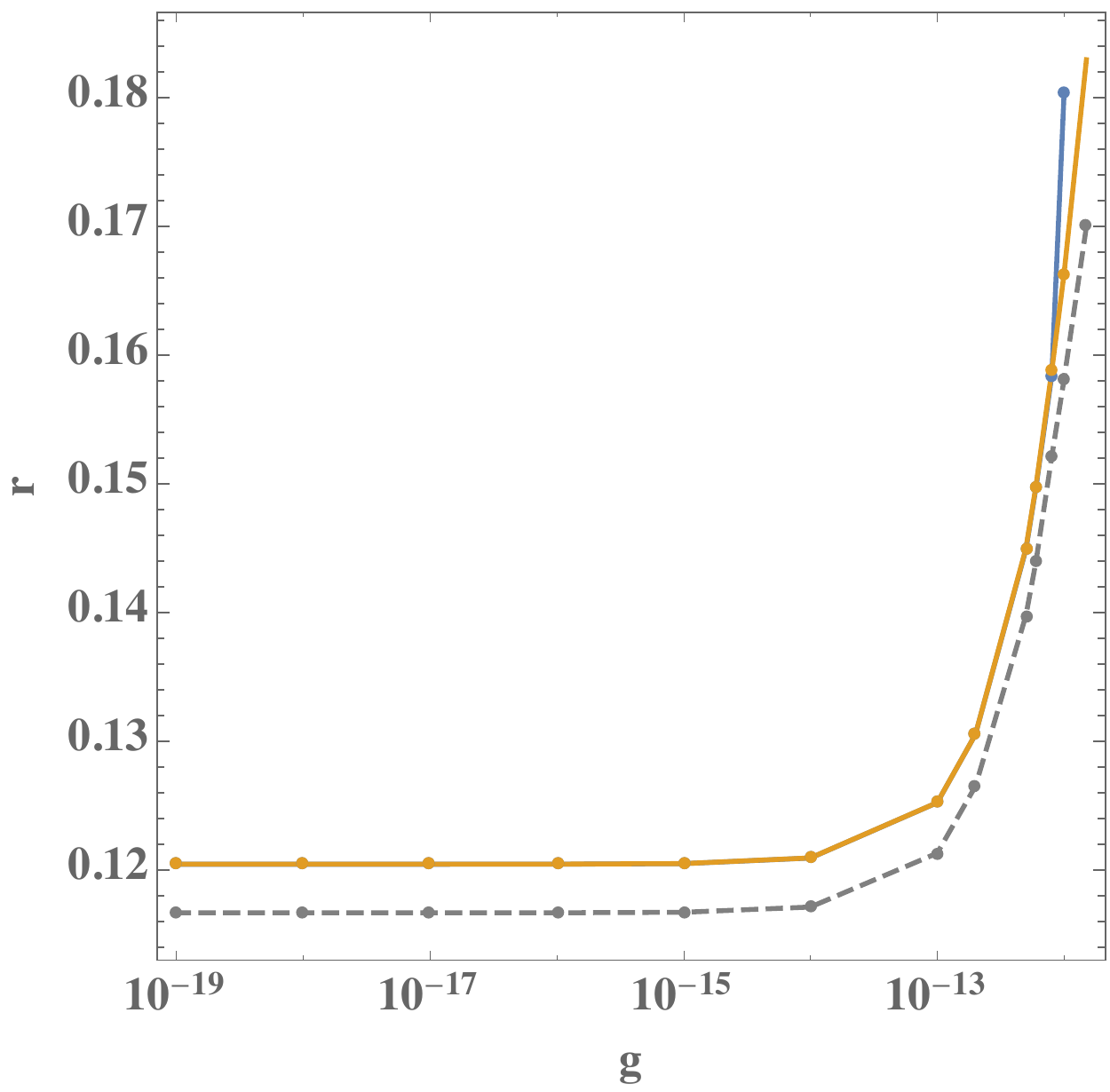}
	\end{minipage}
	\hspace{0.5cm}
	\begin{minipage}[b]{0.3\linewidth}
	\centering
	\includegraphics[width=\textwidth]{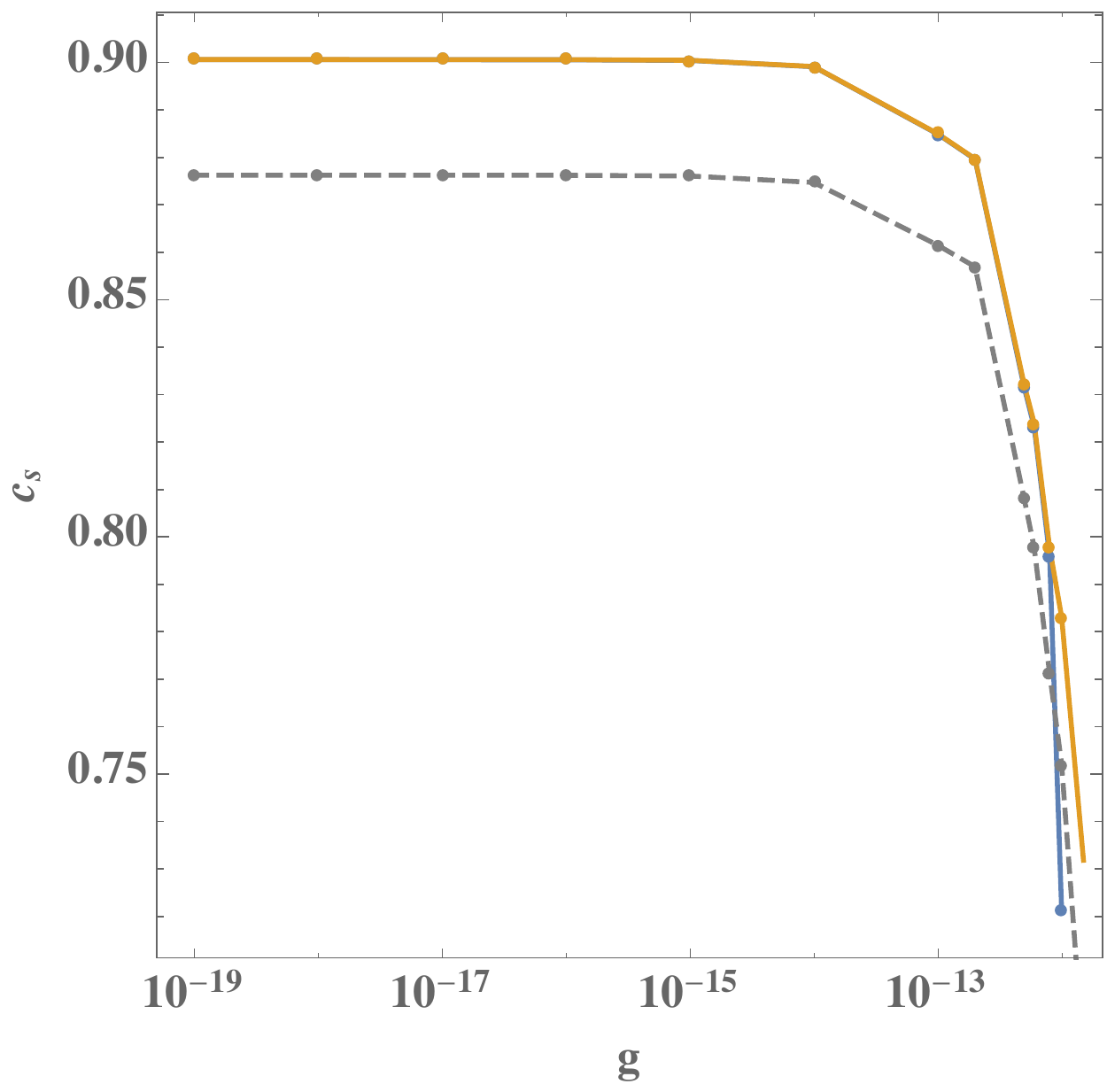}
    \end{minipage}
\caption{$m_L\neq 0$}
	\label{fig:quadratic}
\end{figure}

In order to determine the observable signatures of this type of system and to gauge the accuracy of the perturbative methods developed above it is necessary to perform a numerical analysis of the system. Our strategy is as follows: we start with a parameter set that in the absence of potential coupling ($g=0$) yields $c_s\sim 0.9$, $r\sim 0.12$ and $n_s\sim 0.965$ by setting $\phi_0\ll M_P$ and then increase $g$, while keeping the remaining parameters fixed\footnote{Small adjustments are made to the masses of the heavy and light fields in order to have $\frac{m_H}{H}=10$ and to keep the amplitude of the power spectrum in line with observations for all points.}. Varying $g\in [10^{-19},10^{-12}]$ yields the results displayed in Fig. \ref{fig:quadratic}. We observe that a sizeable reduction in $c_s$ is indeed achievable but that it comes hand in hand with a reddening of $n_s$ and an increase of $r$. These are clear signs that the steepening in the potential is dominating the behaviour of the system.

\paragraph{Regime 2:  $m_L^2\ll g \phi_0^2$\\}

\begin{figure}[h]
	\centering
	\begin{minipage}[b]{0.3\linewidth}
	\centering
	\includegraphics[width=\textwidth]{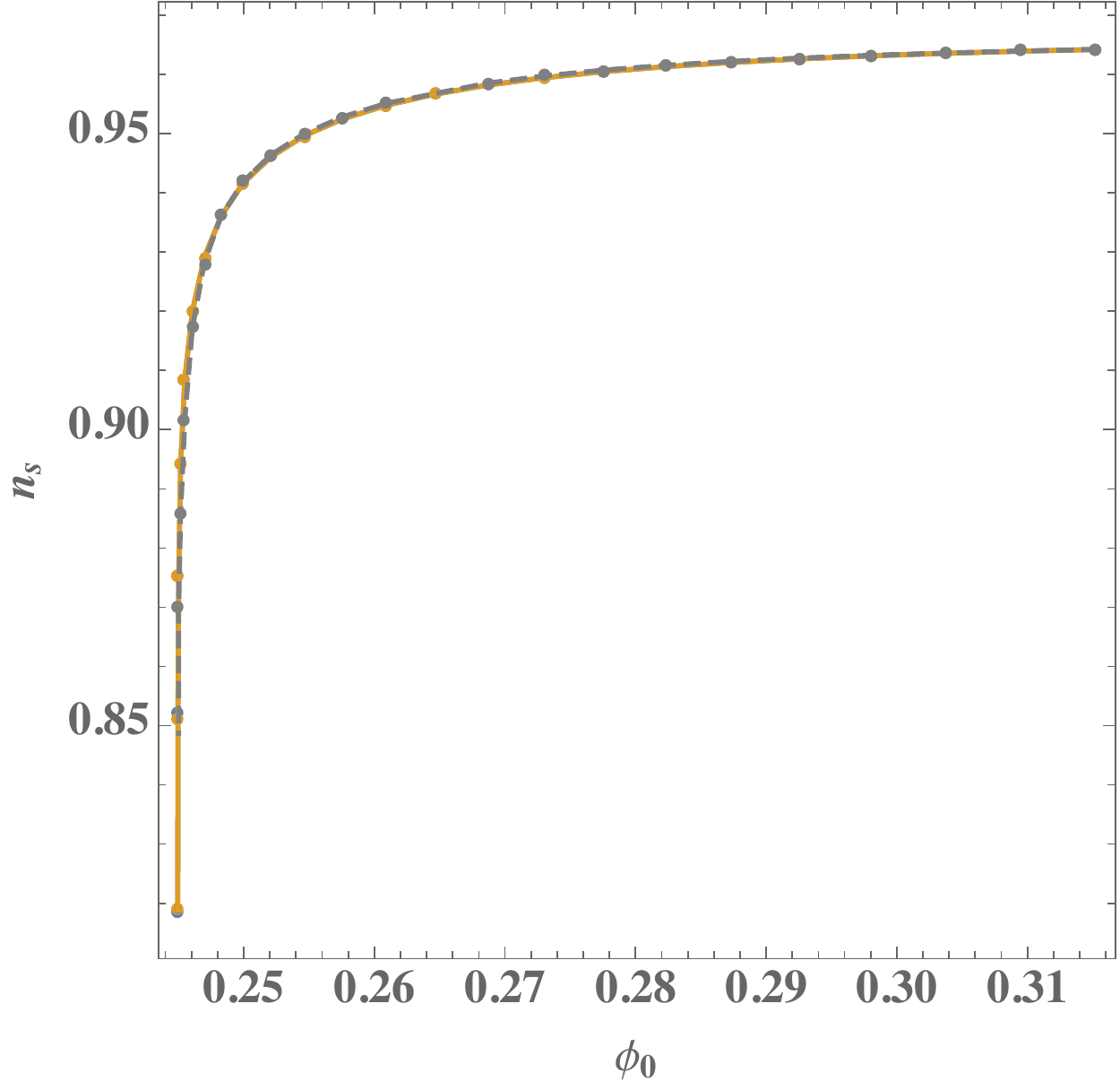}
    \end{minipage}
	\hspace{0.5cm}
	\begin{minipage}[b]{0.29\linewidth}
	\centering
	\includegraphics[width=\textwidth]{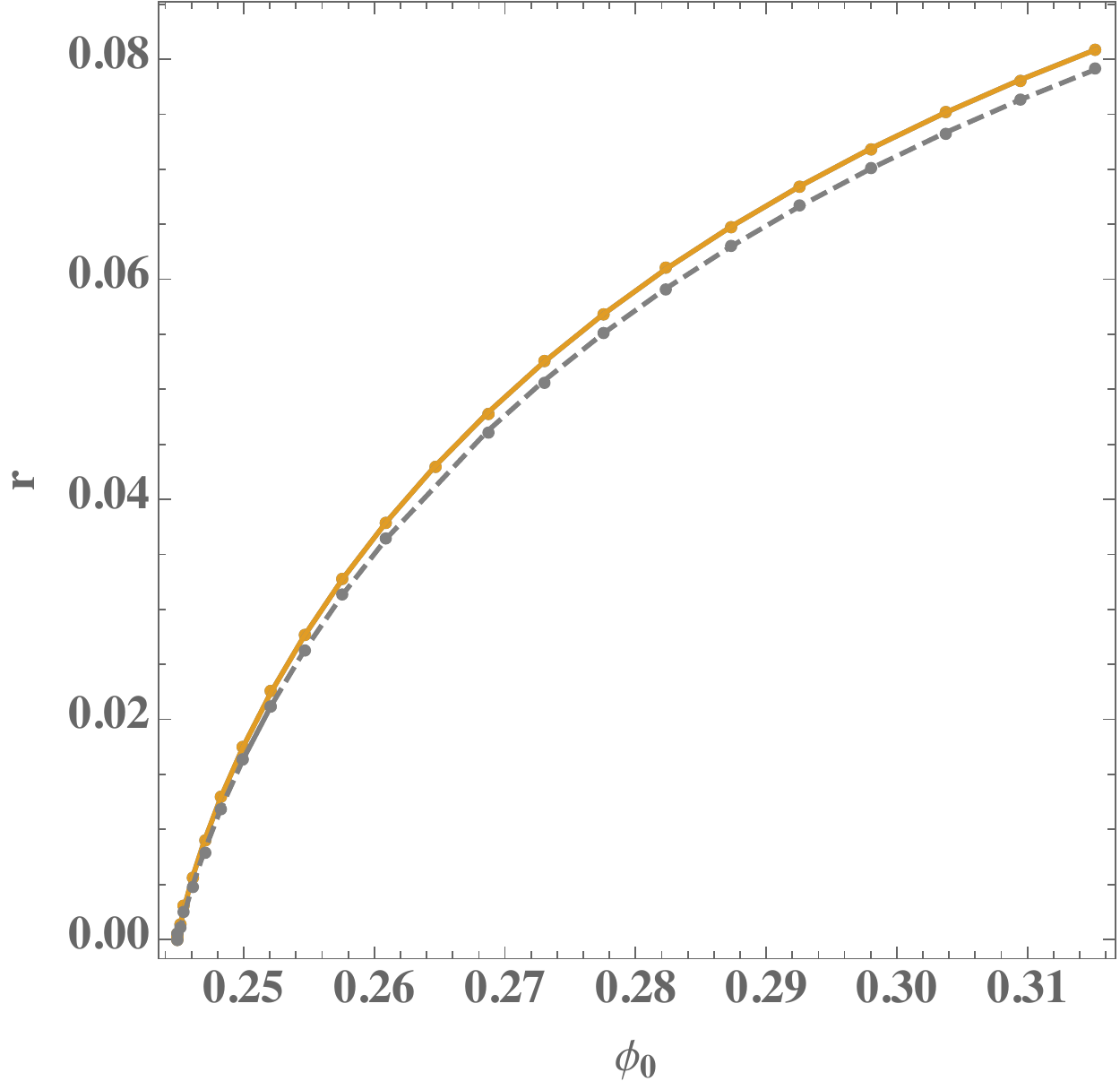}
	\end{minipage}
	\hspace{0.5cm}
	\begin{minipage}[b]{0.29\linewidth}
	\centering
	\includegraphics[width=\textwidth]{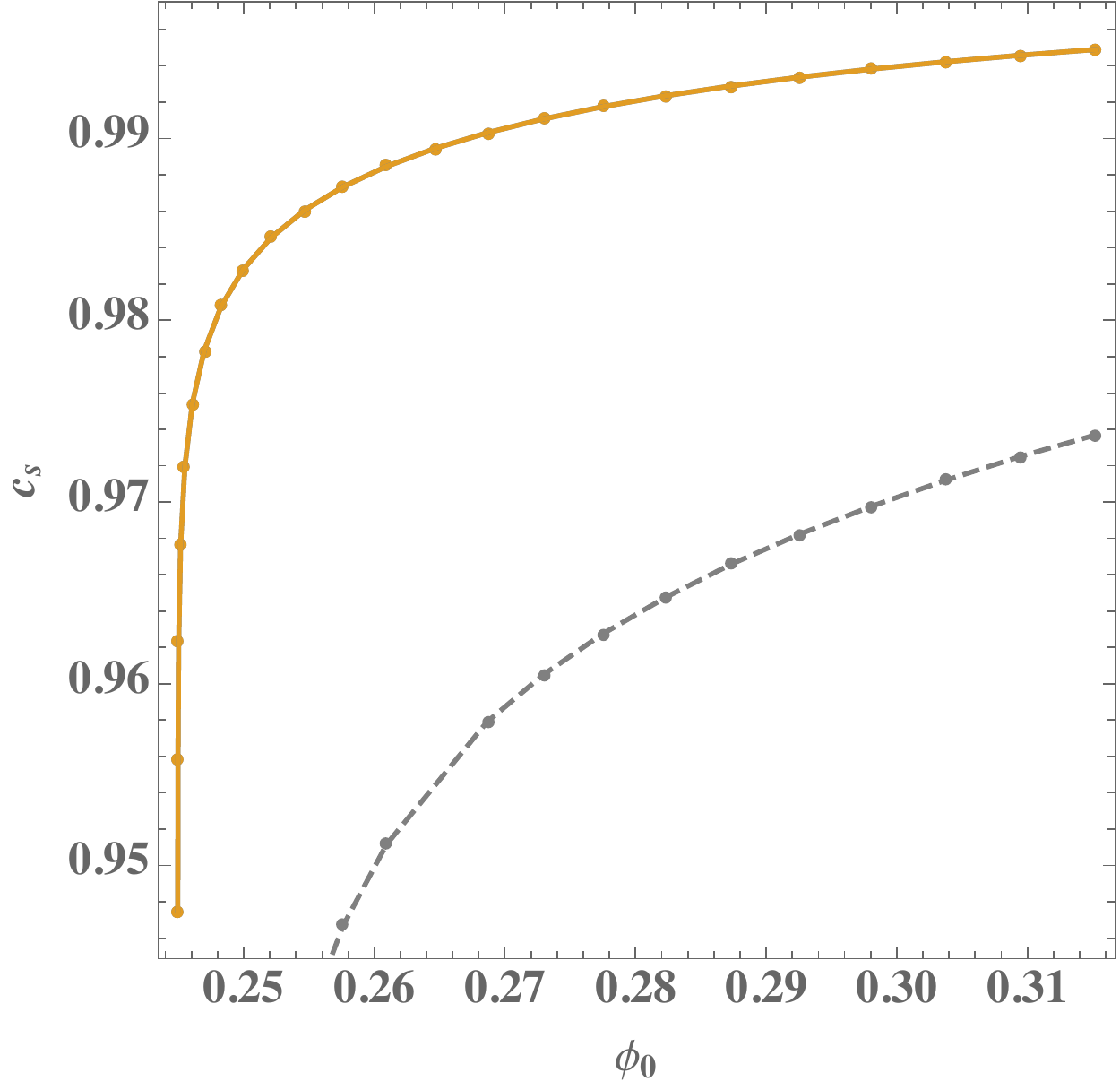}
    \end{minipage}
\caption{Linear $f$ with $m_L= 0$.}
	\label{fig:cosine}
\end{figure}

An alternative  regime of the present example can be found when the inflationary potential is generated by the interaction between the inflaton and the heavy field. This is the case in the stringy constructions of \cite{McAllister:2014mpa} where inflaton self-interactions are often absent, $m_L^2=0$. In this case the scalar potential is reminiscent of that of natural inflation \cite{Freese:1990rb}. For such a regime one can show that the low-energy theory for the canonical field is determined by the two dimensionful parameters $\mu$ and $f_a$ defined in Eq. \eqref{eq:mufa} , where $f_a$ controls the curvature of the potential, while $\mu$ sets its scale. One can then show that 
\be
c_s^{-2}-1\approx \sum_n \frac{\epsilon^n}{ \cos^{2n}(\Phi/f_a)}
\ee
and therefore that reductions of $c_s$ can be achieved if observable inflationary dynamics take place close to the hill-top of the effective potential, where $\Phi/f_a \sim \pi/2$. In this regime however the spectral index deviates significantly from the observed range, as illustrated in Fig. \ref{fig:cosine}. The trajectory in the $(n_s,r)$ plane is identical to that of natural inflation with small ($\mathcal {O}(M_p)$) decay constant, where horizon exit of CMB scales takes place close to the maximum of the cosine potential, and leads to an unacceptably large red tilt. 

\begin{figure}[t!]
	\centering
	\includegraphics[width=0.45\textwidth]{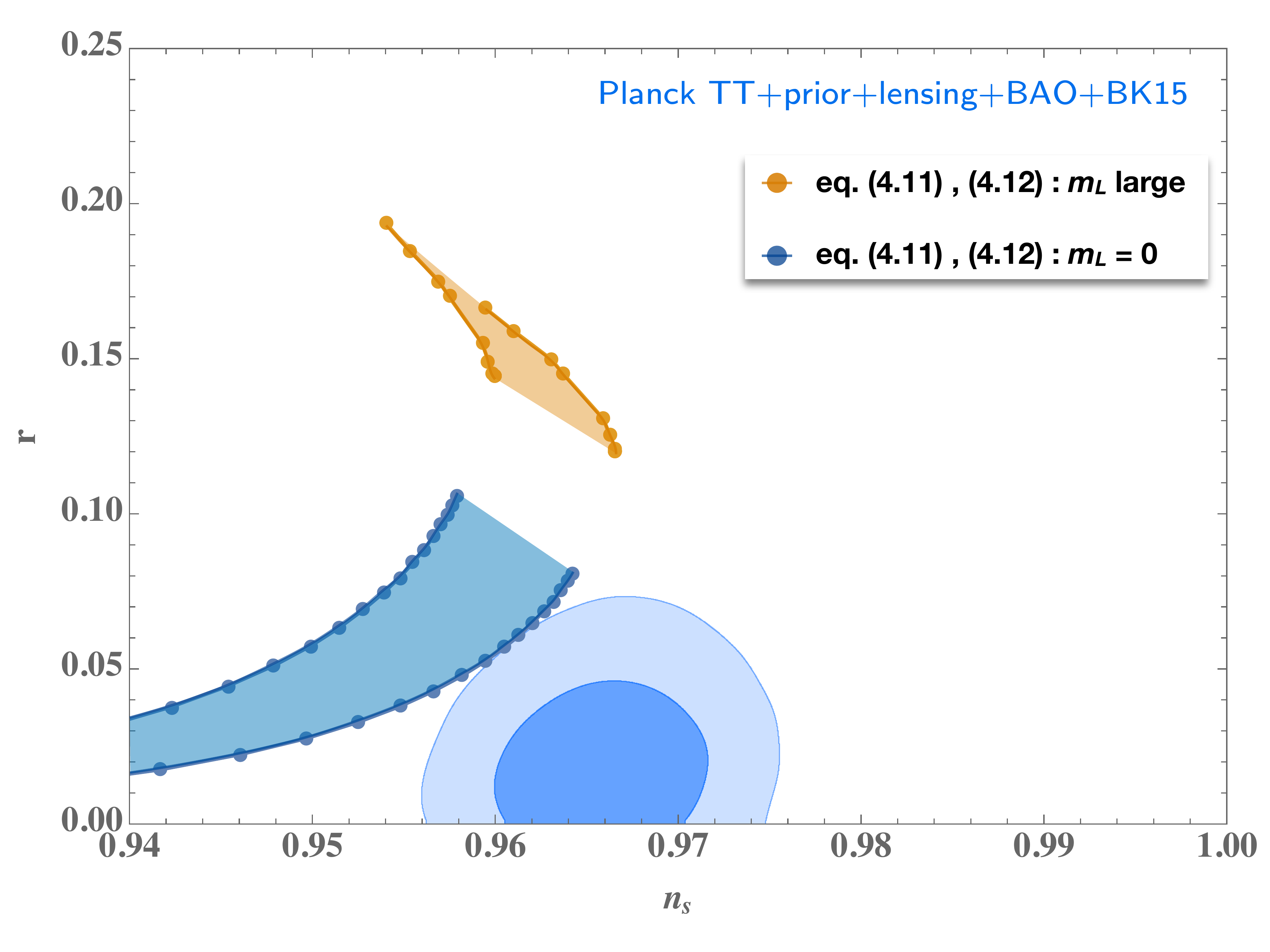}
\caption{Linear $f$ with: $m_L= 0$ (blue-shaded), $m_L$ large (orange-shaded).}
	\label{fig:nsr}
\end{figure}

In Fig. \ref{fig:nsr} we plot the observational signature of the model in both regimes and see that decreasing $c_s$ does not help in bringing the model into the observationally allowed region, in fact the opposite holds true: reductions of $c_s$ come hand in hand with steepening effects that push the spectrum away from scale invariance.

\subsection{Exponential kinetic coupling}

As a final example in this section we consider $f(\phi_H)=e^{\phi_H/\Lambda}$. Unlike in the previous section  is not possible to analytically find the canonical variable, $\Phi$, which renders the analysis more involved. We can however perform an approximate canonical normalisation that allows us to probe the system in two different regimes, where analytical tools can then be used. 

Applying the method described at the start of this section one finds that interaction between the light and heavy fields induces a subluminal speed of sound for the scalar perturbations that is given by
\be
c_s^{-2}=1+ \sum_{n=1}^6 a_{2n} \left(\frac{1}{\Lambda^2 (m_H^2+g \phi_L^2(\Phi))}\right)^n \dot{\Phi}^{2n}
\label{eq:csExp}
\ee 
where both EFTs predict
\be
a_2=  4\ , \ a_4= 16 \ ,\ a_6=72 \ , \ a_8=1024 \ ,\ a_{10}=5000 \ ,
\ee
differing only twelfth order in $\dot{\phi}_L$.

Neglecting overall numerical coefficients, one may use the background eoms to write
\be
c_s^{-2}-1=\sum_{n\ge1} \epsilon^n \left(\frac{H}{M_H}\right)^{2n}\left(\frac{M_P}{\Lambda}\right)^{2n}\ ,
\ee
where we defined the total mass of the heavy field during inflation as $M_H^2\equiv m_H^2+g \phi_L^2$. It then becomes evident that reductions of the speed of sound can be obtained in the small $\Lambda$ regime, just like in the exponential example of Sec. 3.

In what follows we study the dynamics of the system in two different regimes, a strong mixing regime, where the canonical field $\Phi$ approximately can be expressed as 
 \be
 \Phi\approx\phi_L-\frac{\phi_0 m_H^2}{\Lambda g \phi_L} \qquad\text{for}\quad  \frac{g \phi_L^2}{m_H^2}\gg1  \quad\text{and}\quad   \frac{\Lambda}{\phi_0} \gg  \frac{m_H^2}{g \phi_L^2}
 \label{eq:PhiStrong}
 \ee
 and a weak mixing regime where it is instead given by 
 \be
\Phi\approx e^{\phi_0/\Lambda} \phi_L\left(1-\frac{1}{3}\frac{\phi_0}{\Lambda}\frac{g \phi_L^2}{ m_H^2}\right) \qquad\text{for}\quad  \frac{g \phi_L^2}{m_H^2}\ll1  \quad\text{and}\quad   \frac{\Lambda}{\phi_0} \gg \frac{g \phi_L^2}{m_H^2}\ .
 \ee
In both regimes we will consider the case when the inflationary potential is dominated by the inflaton's self interactions (regime 1) and when it is predominantly generated by interaction with the heavy field (regime 2). 

\paragraph{Regime 1: $m_L^2\gg g \phi_0^2$\\}
In this regime the inflationary potential comes predominantly from the inflaton's self-interactions. We therefore have $H^2\propto m_L^2$, in particular $H$ is  essentially independent of the post inflationary vev of the heavy field $\phi_0$ that, as we will see below, parametrises the size of the corrections (both coming from kinetic interactions and mixing in the potential) to the starting quadratic potential for the inflaton.

\begin{itemize}
\item $\frac{g \phi_L^2}{m_H^2}\gg1$\\

The potential for the approximate canonical variable $\Phi$ can be written as a series in $\frac{m_H^2}{g \Phi^2}$, whose first terms are

\be
V_{eff}(\Phi)=\frac{m_L^2}{2}\Phi^2\left( 1+\frac{\phi_0 m_H^2}{\Lambda g \Phi^2}\right)^2 + \frac{m_H^2 \phi_0^2}{2}\left(1-\frac{m_H^2}{g \Phi^2}\left(1-2 \frac{\phi_0}{\Lambda}\frac{m_H^2}{g \Phi^2}-\frac{m_H^2}{g \Phi^2}\right)\right)\ .
\ee
In the limit of small $\Lambda$, where reductions of $c_s$ are in principle attainable, the effective potential induces a mild reddening of $n_s$, and an increase of the tensor fraction. Since the reduction in $c_s$ is rather modest, the potential is the determining factor in the behaviour of the observables displayed in Fig. \ref{fig:ExpmLStrong}. Indeed the lower half of the $\phi_0/\Lambda$ range in Fig. \ref{fig:ExpmLStrong} is be well described by a two derivative action with a potential of the form $V\sim \Phi^2(1+\alpha/\Phi^2)$ with $\alpha\ll1$.

\begin{figure}[t!]
		\centering
	\begin{minipage}[b]{0.3\linewidth}
	\centering
	\includegraphics[width=\textwidth]{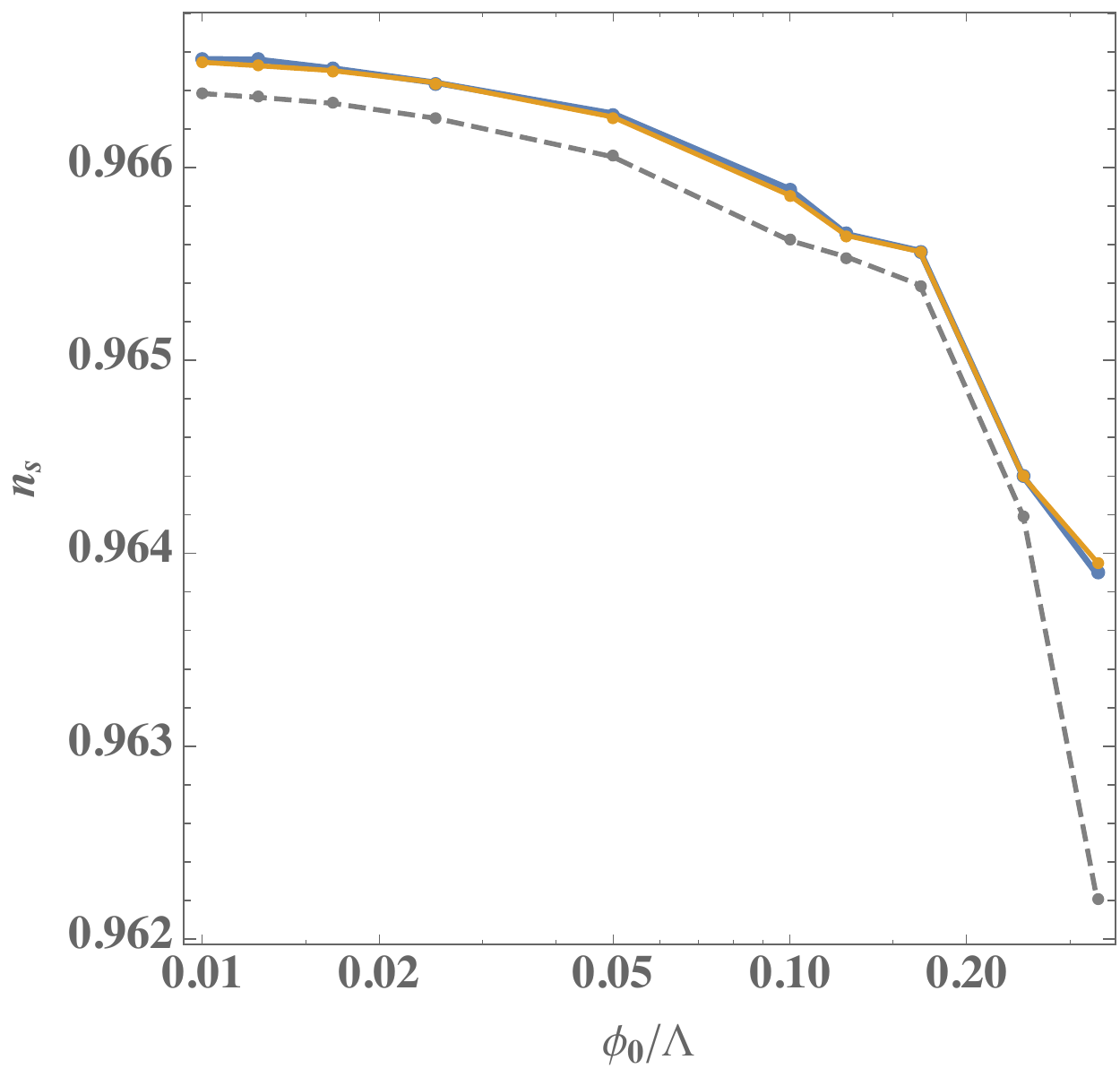}
    \end{minipage}
	\hspace{0.5cm}
	\begin{minipage}[b]{0.29\linewidth}
	\centering
	\includegraphics[width=\textwidth]{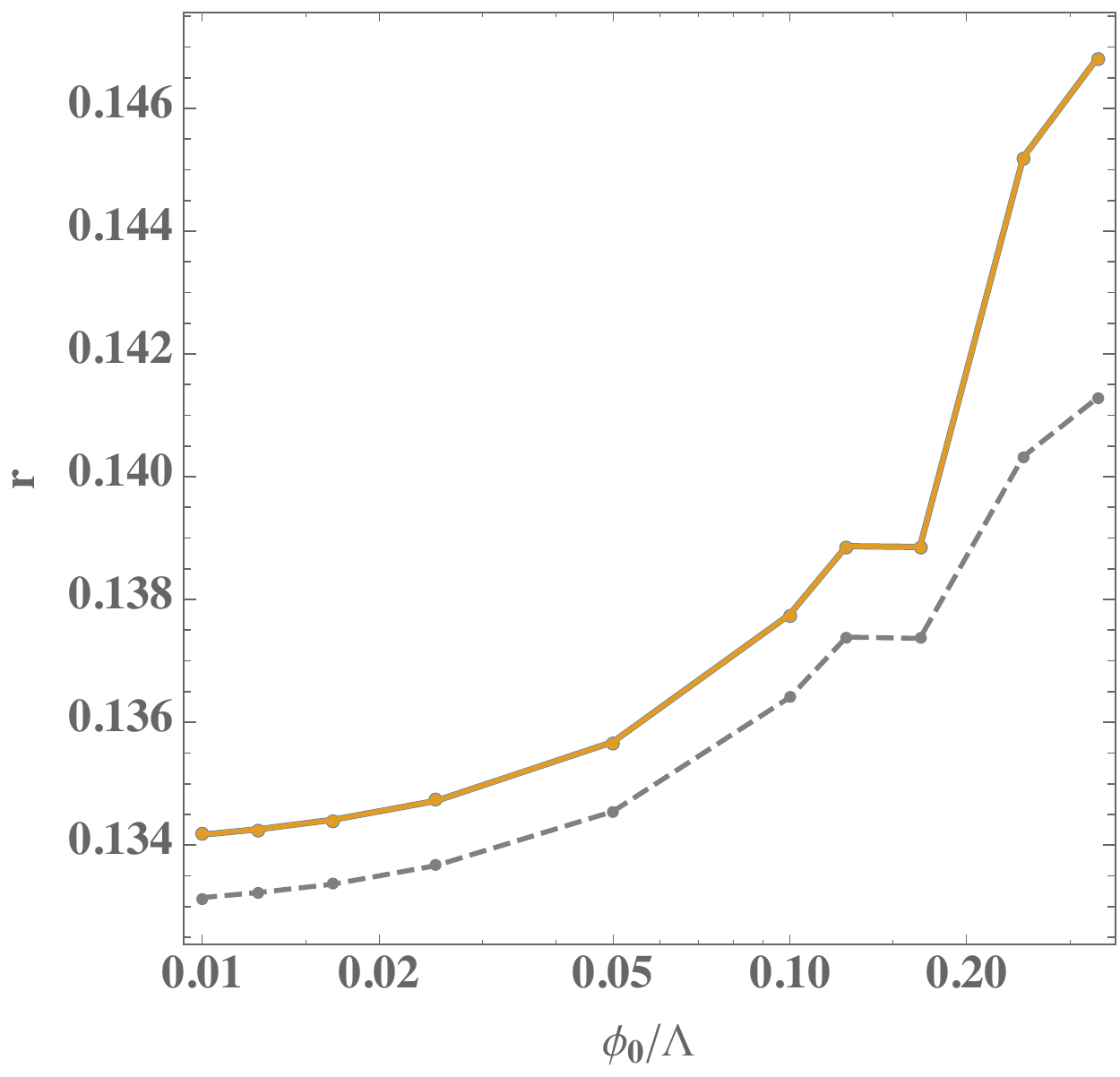}
	\end{minipage}
	\hspace{0.5cm}
	\begin{minipage}[b]{0.29\linewidth}
	\centering
	\includegraphics[width=\textwidth]{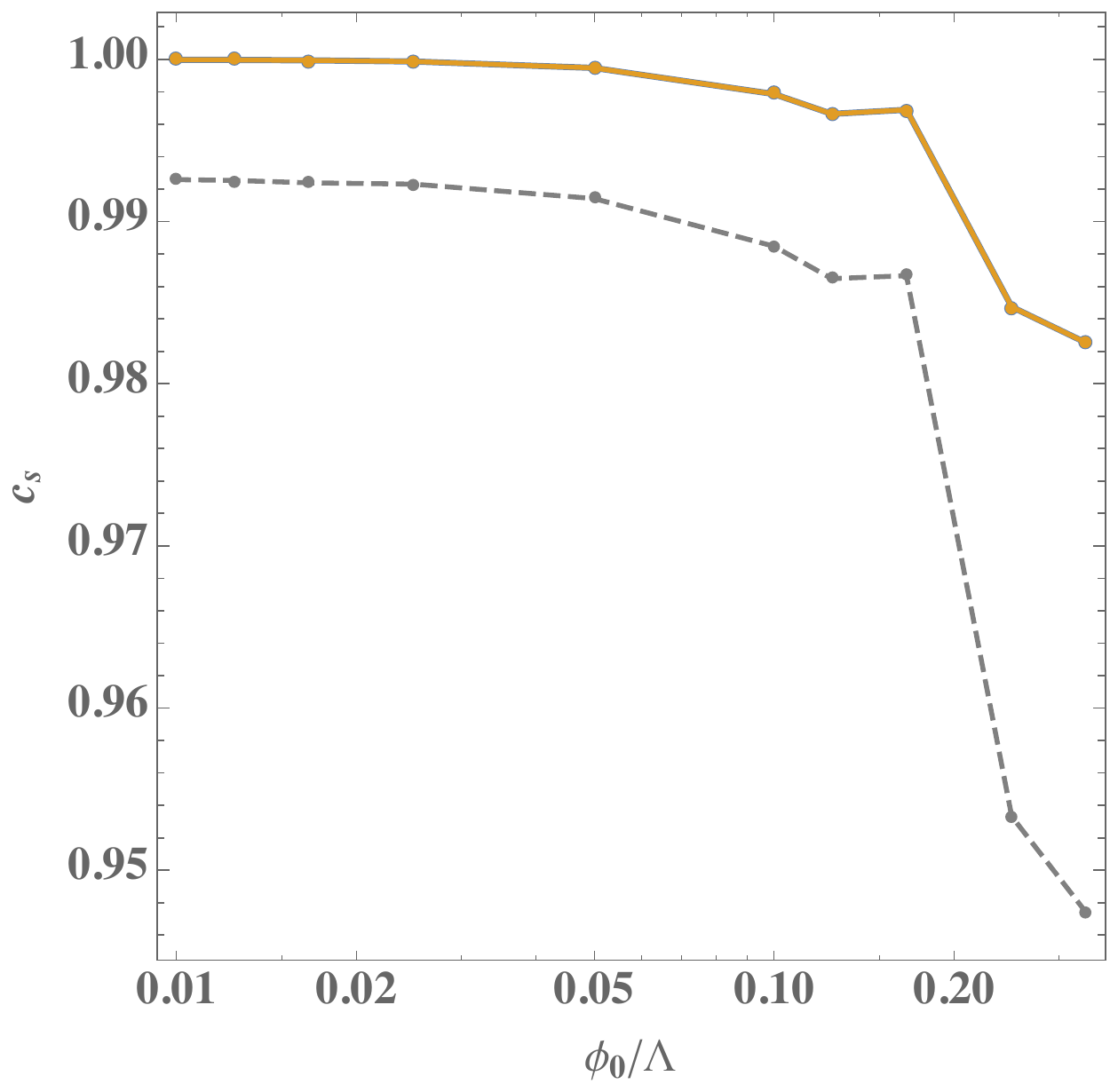}
        \end{minipage}
\caption{Observables for $f=e^{\phi_H/\Lambda}$ in the regime $m_L\gg g \phi_0^2$ and $\frac{g \phi_L^2}{m_H^2}\gg1$.} 
	\label{fig:ExpmLStrong}
\end{figure}

\item $\frac{g \phi_L^2}{m_H^2}\ll1$\\

In this case the potential admits an expansion in powers of $\frac{g \Phi^2}{m_H^2}$ of the form 
\be
V_{eff}(\Phi)=\frac{\mu_1^2}{2}\Phi^2\left( 1+\frac{\phi_0 g }{3 \Lambda m_H^2}e^{-2 \phi_0/\Lambda}  \Phi^2 \right)^2+\frac{\mu_2^2}{2}\Phi^2\left(1- \frac{\mu_2^2 \Phi^2}{ m_H^2 \phi_0^2}+\frac{ 2 \mu_2^2 \Phi^2}{3 m_H^2 \phi_0^2}\frac{\phi_0}{\Lambda}\right)
\ee
where the effective mass parameters are defined as $\mu_1^2\equiv m_L^2 e^{-2 \phi_0/\Lambda}$ and  $\mu_2^2\equiv g \phi_0^2 e^{-2 \phi_0/\Lambda}$.
Noting that corrections to the quadratic potential for the inflaton (both coming from kinetic interactions and mixing in the potential)  are proportional to $e^{-\phi_0/\Lambda}$,  and that deviations from $c_s=1$ are attainable for small $M_P/\Lambda$, one can tune down the speed of sound by scanning over the large $\phi_0/\Lambda$ part of parameter space. This way we can reduce $c_s$, and consequently $r$, without causing significant shift of $n_s$ form the observed value.  Numerical results are depicted in Fig. \ref{fig:ExpmLWeak} and are identical to those of Sec. \ref{sec:exp1}: a reduction of $c_s$ lowers $r$ while the small value of $\phi_0$ ensures that the spectral index remains essentially constant. 
\begin{figure}[h]
		\centering
	\begin{minipage}[b]{0.3\linewidth}
	\centering
	\includegraphics[width=\textwidth]{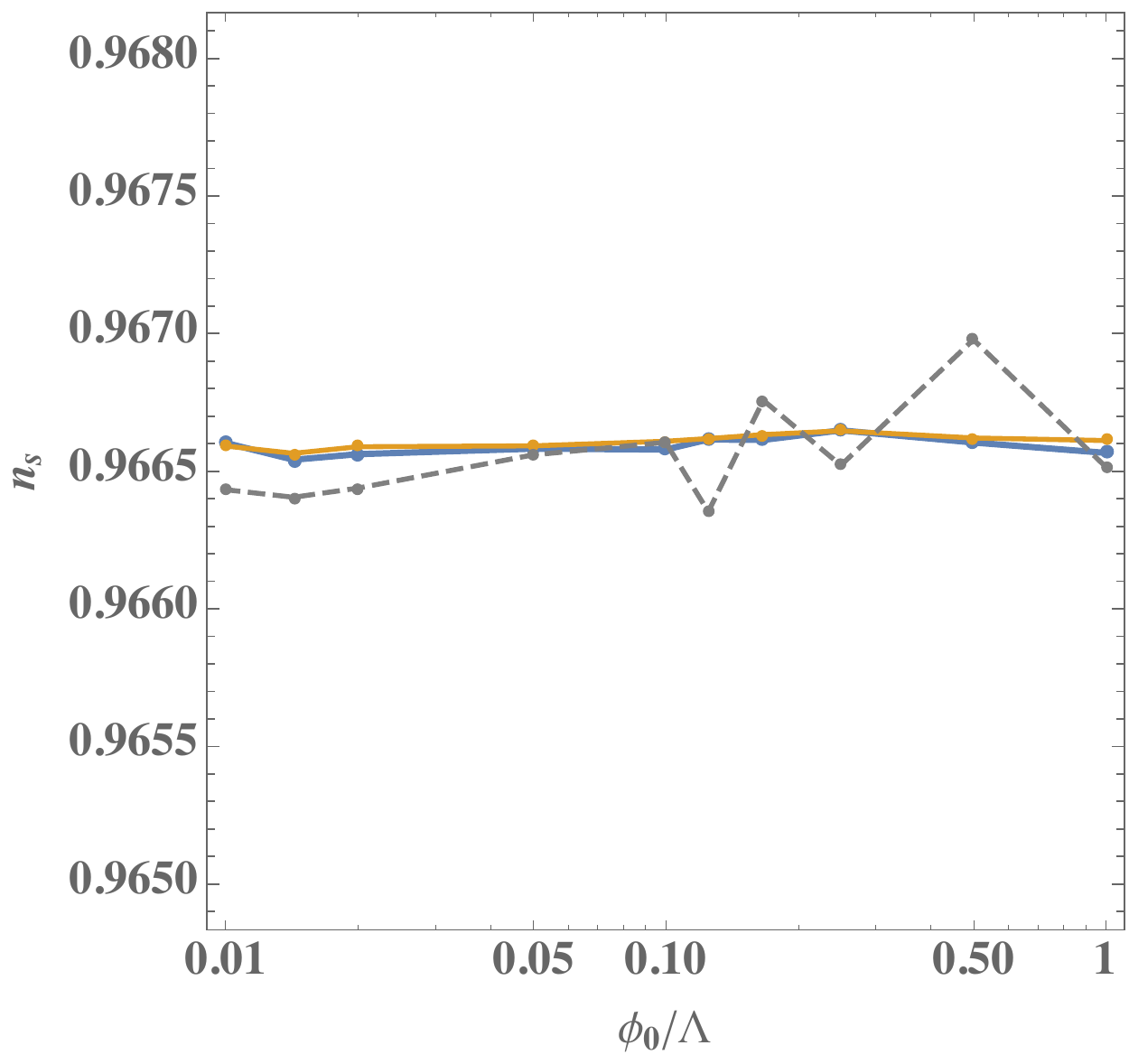}
    \end{minipage}
	\hspace{0.5cm}
	\begin{minipage}[b]{0.29\linewidth}
	\centering
	\includegraphics[width=\textwidth]{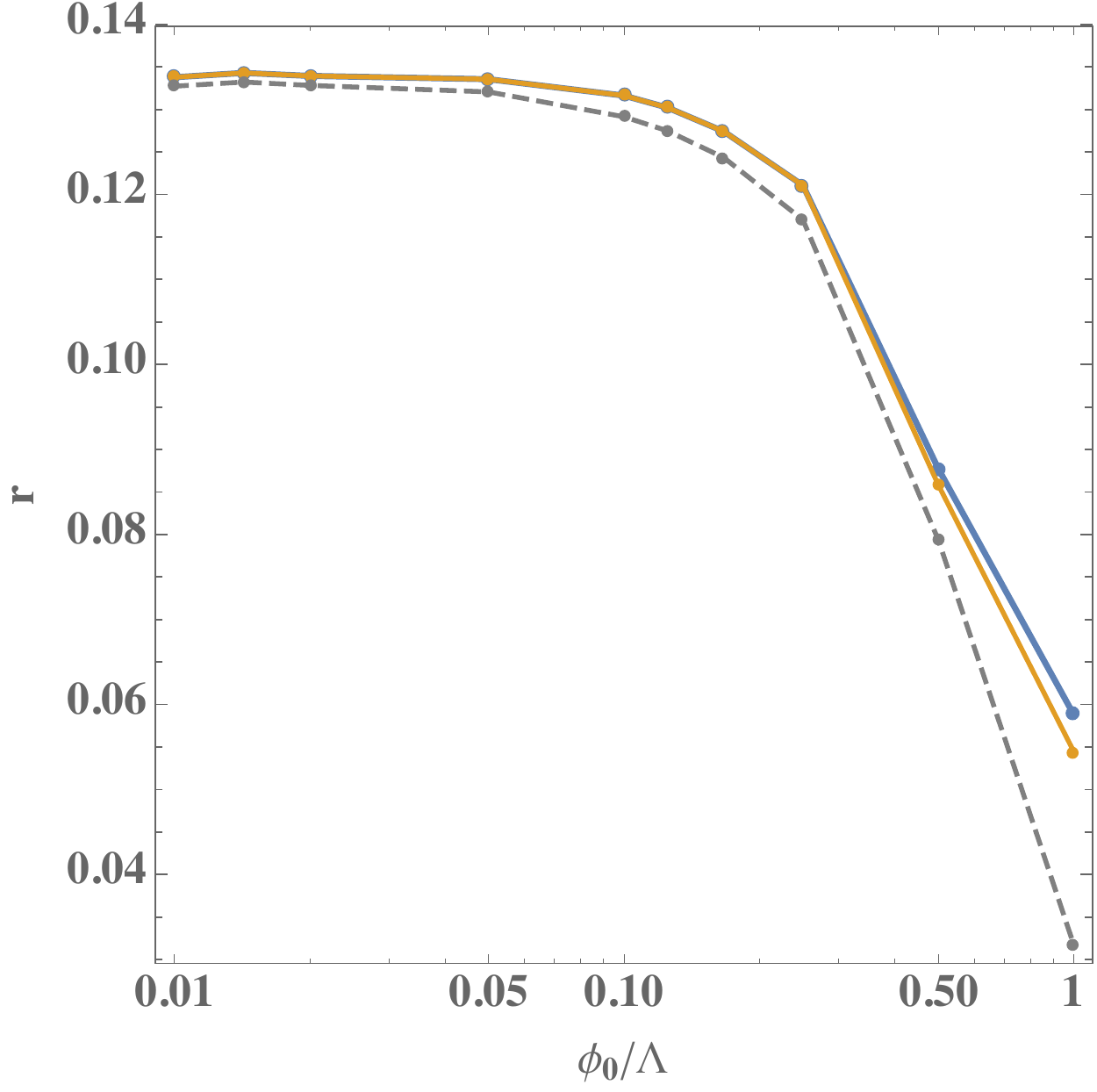}
	\end{minipage}
	\hspace{0.5cm}
	\begin{minipage}[b]{0.29\linewidth}
	\centering
	\includegraphics[width=\textwidth]{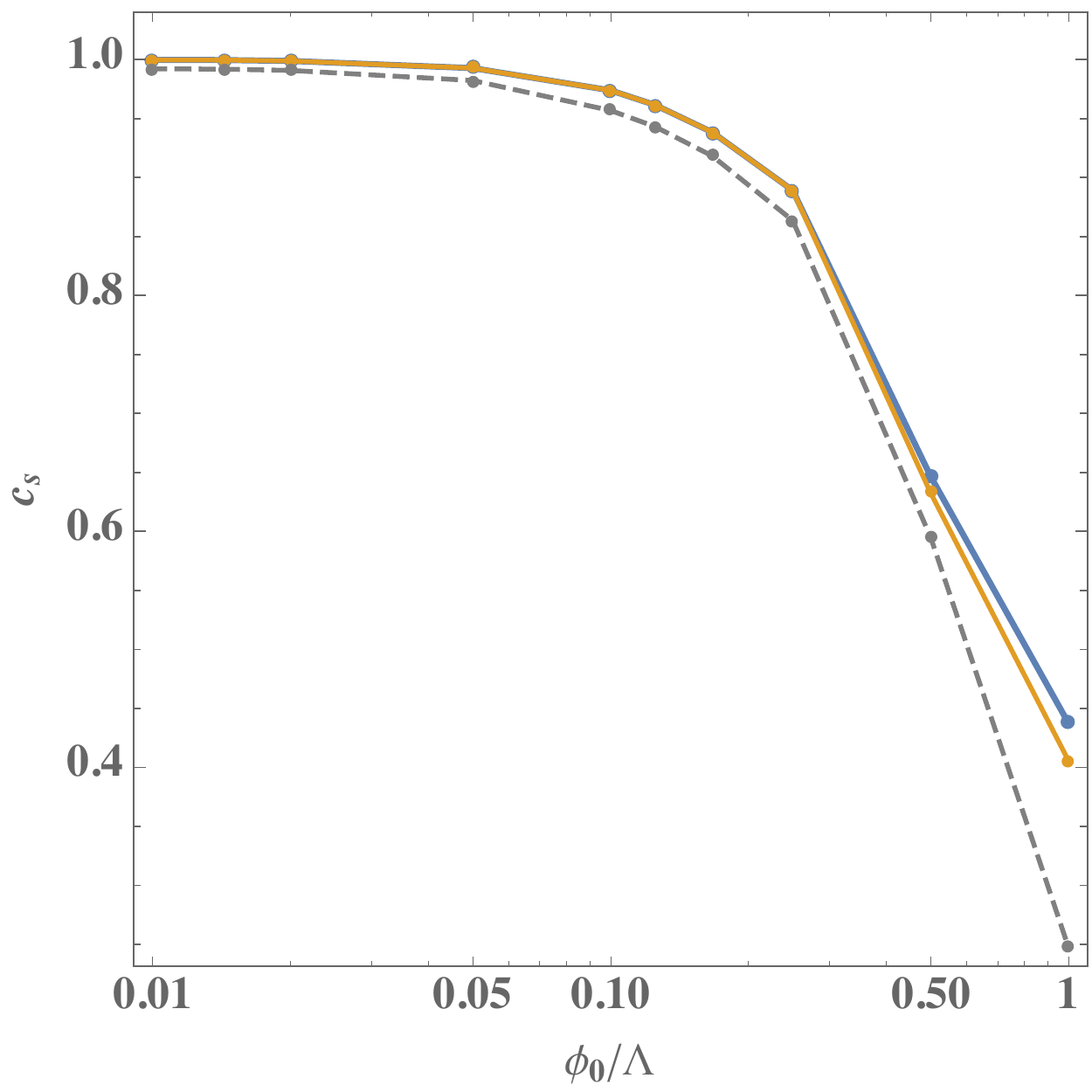}
        \end{minipage}
\caption{Observables for $f=e^{\phi_H/\Lambda}$ in the regime $m_L\gg g \phi_0^2$ and $\frac{g \phi_L^2}{m_H^2}\ll1$.} 
	\label{fig:ExpmLWeak}
\end{figure}

\end{itemize}

\paragraph{Regime 2: $m_L^2\ll g \phi_0^2$\\}
Let us now consider the possibility that the inflationary potential is generated by the interaction between the inflaton and the heavy field and consider a strong mixing/flattening  $\frac{g \phi_L^2}{m_H^2}\gg1$ and a weak mixing/flattening $\frac{g \phi_L^2}{m_H^2}\ll1$ regimes. For simplicity we will set $m_L=0$ in what follows. In this region of parameter space $H^2\propto \phi_0^2$ and so $\phi_0$ sets the inflationary scale and is therefore constrained by observations (unlike in regime 1), in particular by the amplitude of the scalar power spectrum.

\begin{figure}[t!]
	\centering
	\begin{minipage}[b]{0.30\linewidth}
	\centering
	\includegraphics[width=\textwidth]{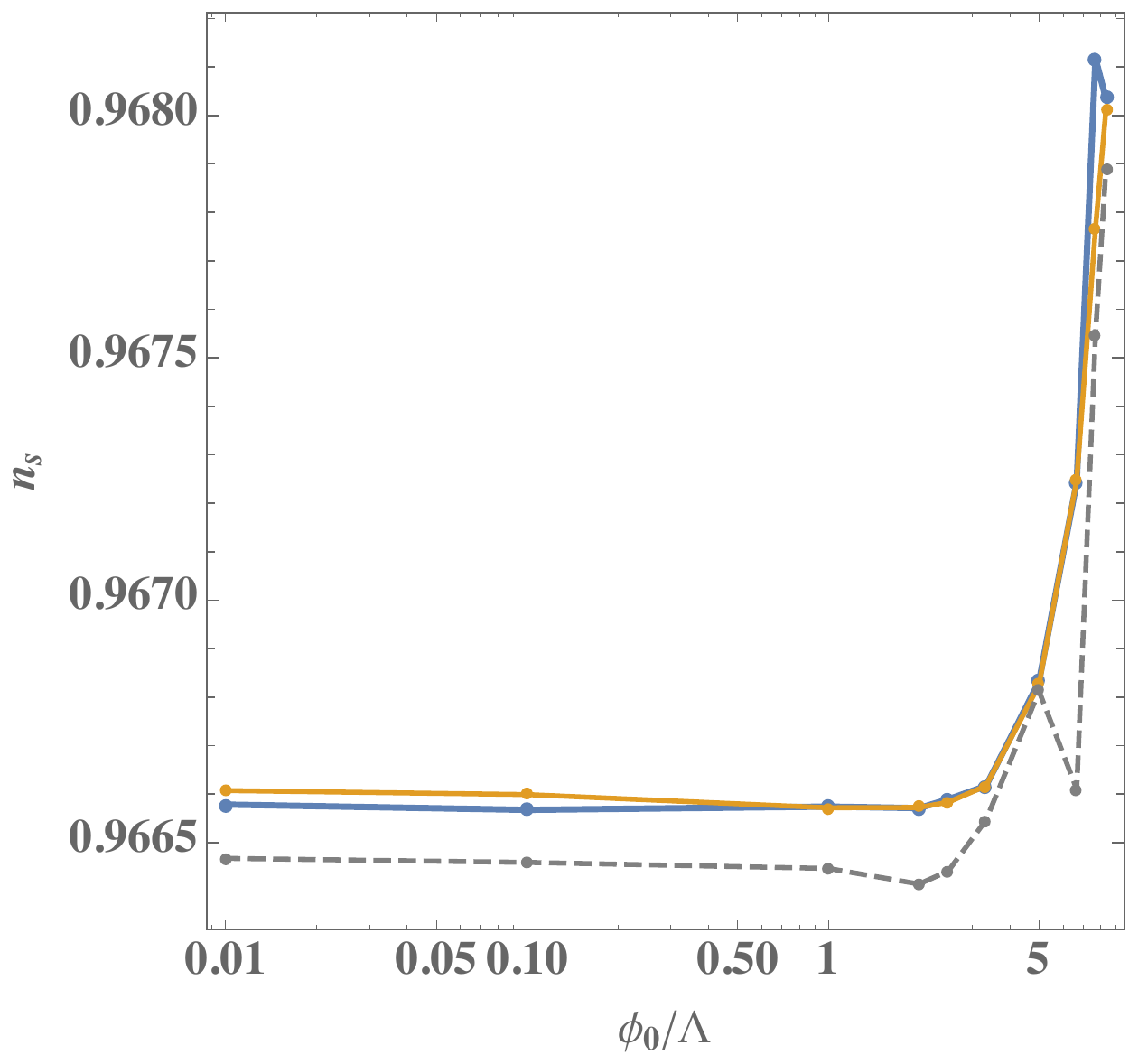}
    \end{minipage}
	\hspace{0.5cm}
	\begin{minipage}[b]{0.29\linewidth}
	\centering
	\includegraphics[width=\textwidth]{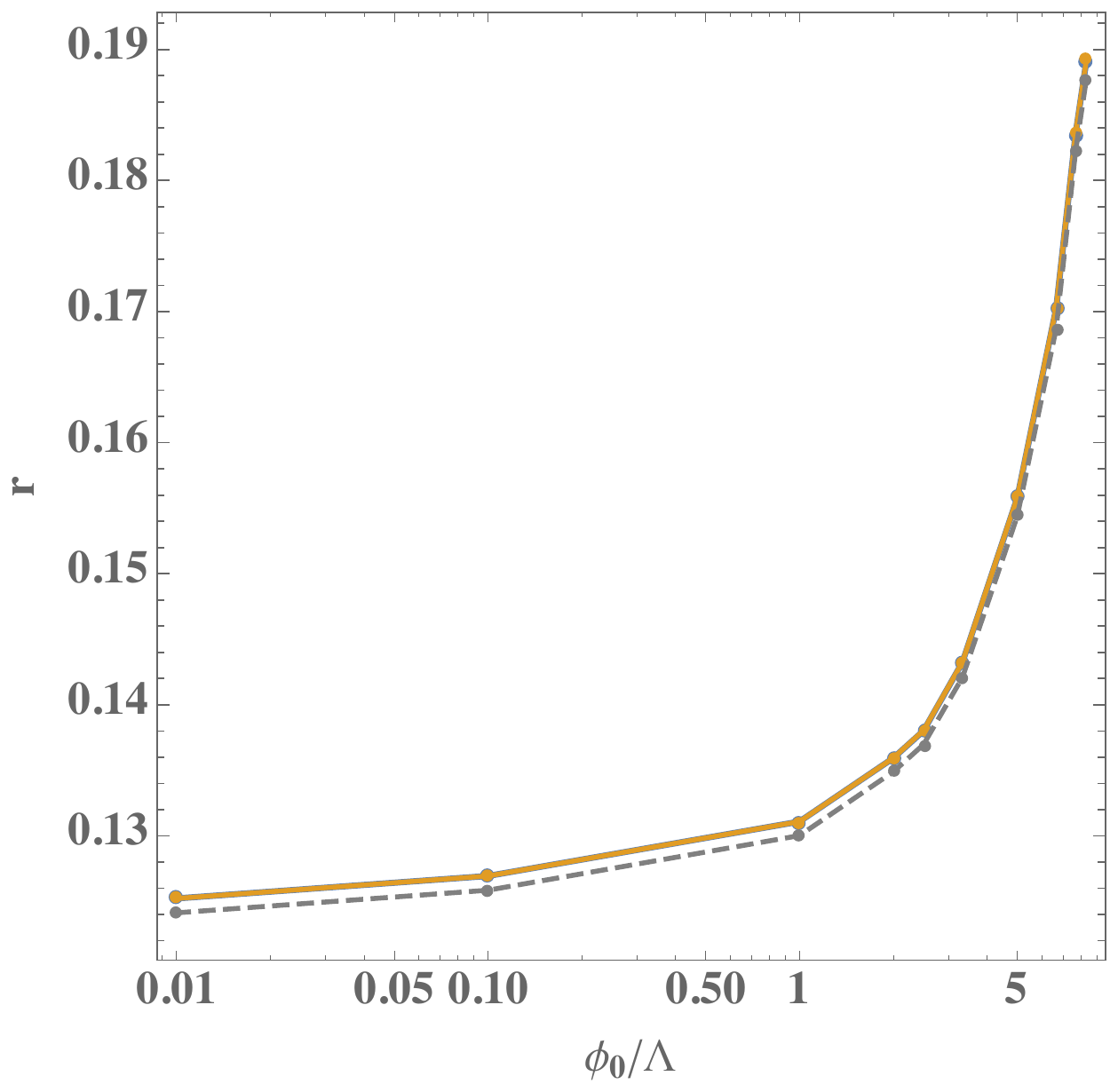}
	\end{minipage}
	\hspace{0.5cm}
	\begin{minipage}[b]{0.29\linewidth}
	\centering
	\includegraphics[width=\textwidth]{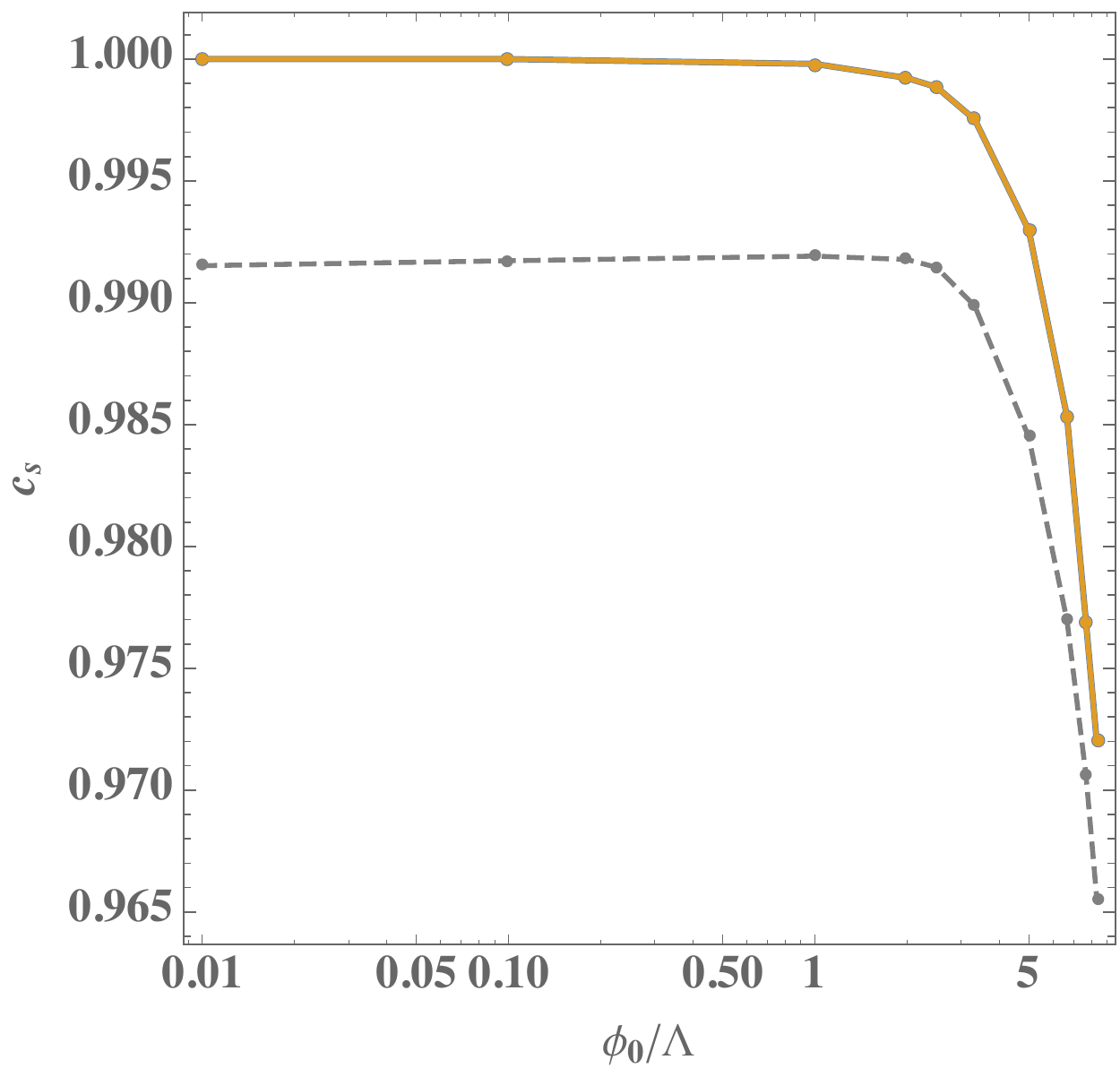}
    \end{minipage}
\caption{Observables for $f=e^{\phi_H/\Lambda}$ in the regime $m_L^2\ll g \phi_0^2$ and $ \frac{g \phi_L^2}{m_H^2}\ll1$.}
	\label{fig:ExpNonFlat}
\end{figure}

\begin{itemize}
\item $\frac{g \phi_L^2}{m_H^2}\gg1$\\

 This regime corresponds to having a strong flattening effect in the scalar potential, which gets strongly distorted from its original quadratic form and in the inflationary region is now given by
 \be
V=\frac{m_H^2 \phi_0^2}{2}\left(1-\frac{m_H^2}{g \Phi^2}\left(1-2 \frac{\phi_0}{\Lambda}\frac{m_H^2}{g \Phi^2}-\frac{m_H^2}{g \Phi^2}\right)\right)\ .
\ee
Noting that $M_H^2\approx g \Phi^2$ in this regime, one can also write the speed of sound as
\be
c_s^2-1\approx \sum_{n\ge1} \epsilon^n \left(\frac{\phi_0}{\Lambda}\right)^{2n}\left(\frac{m_H^2}{g \Phi^2}\right)^n
\label{eq:cs22}
\ee
where we are omitting numerical coefficients. The last factor in Eq. \eqref{eq:cs22} is hierarchically small, c.f. Eq.  \eqref{eq:PhiStrong}, and therefore reductions of $c_s$ are only attainable in the large $\phi_0/\Lambda$ regime. Results are displayed in Fig. \ref{fig:ExpFlat}. The scaling of $n_s$ and $r$ with $\phi_0/\Lambda$ taken in conjunction with the negligible shift in $c_s$ indicate that the HD operators are unimportant in this regime and that observables are dominated by the higher order $\frac{\phi_0}{\Lambda}\frac{m_H^2}{g \Phi^2}$ corrections to the scalar potential, these render it steeper and cause an increase of $r$ and an abrupt  departure from scale invariance. Note that the results from the full two-field computation are absent from Fig. \ref{fig:ExpFlat}, this is due to the fact that in this regime, performing such a computation to the desired level of numerical precision becomes computationally very costly with the tools we are using, thereby demonstrating the usefulness of the EFT methods developed above.

\begin{figure}[h!]
	\centering
	\begin{minipage}[b]{0.285\linewidth}
	\centering
	\includegraphics[width=\textwidth]{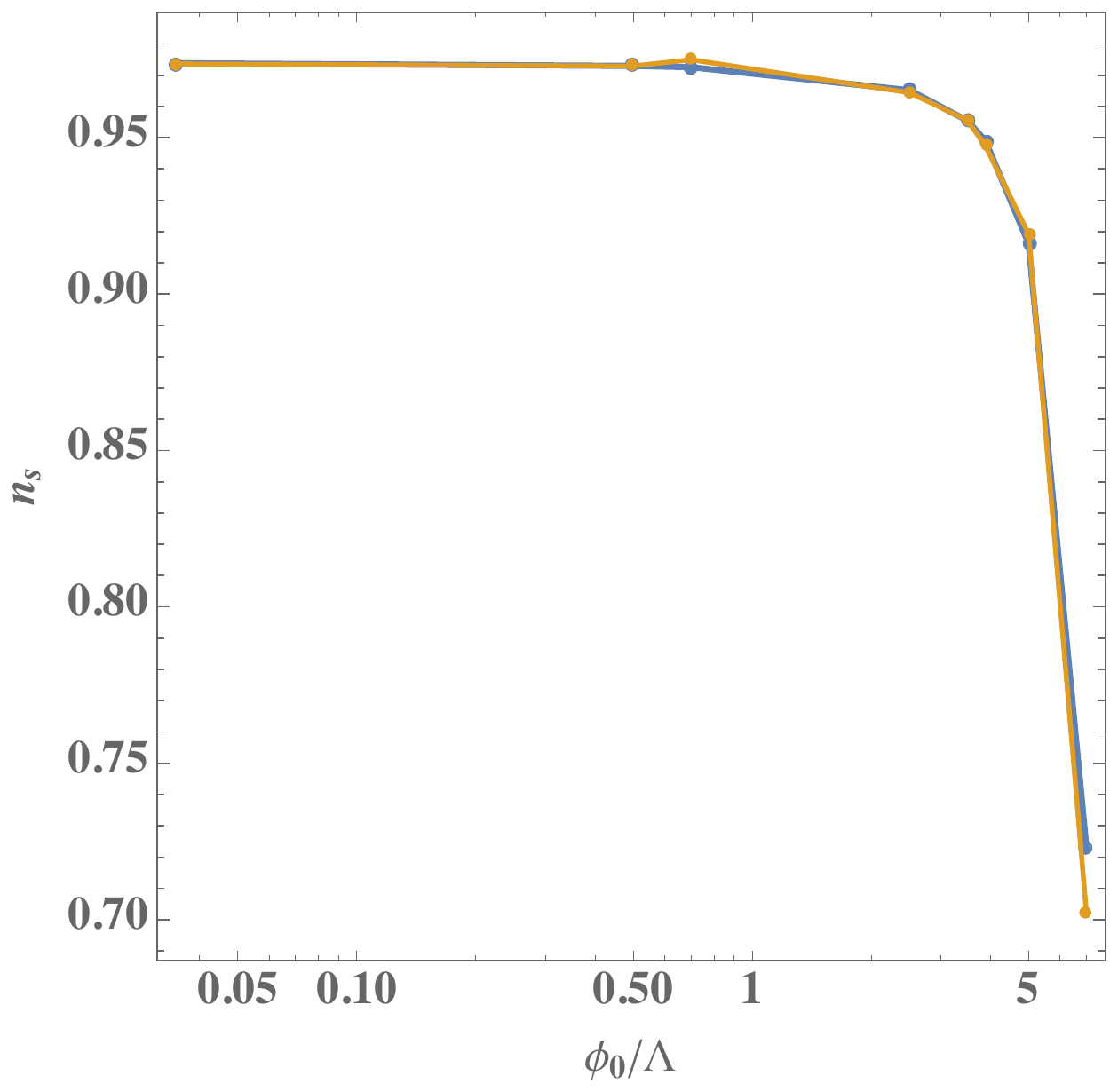}
    \end{minipage}
	\hspace{0.5cm}
	\begin{minipage}[b]{0.3\linewidth}
	\centering
	\includegraphics[width=\textwidth]{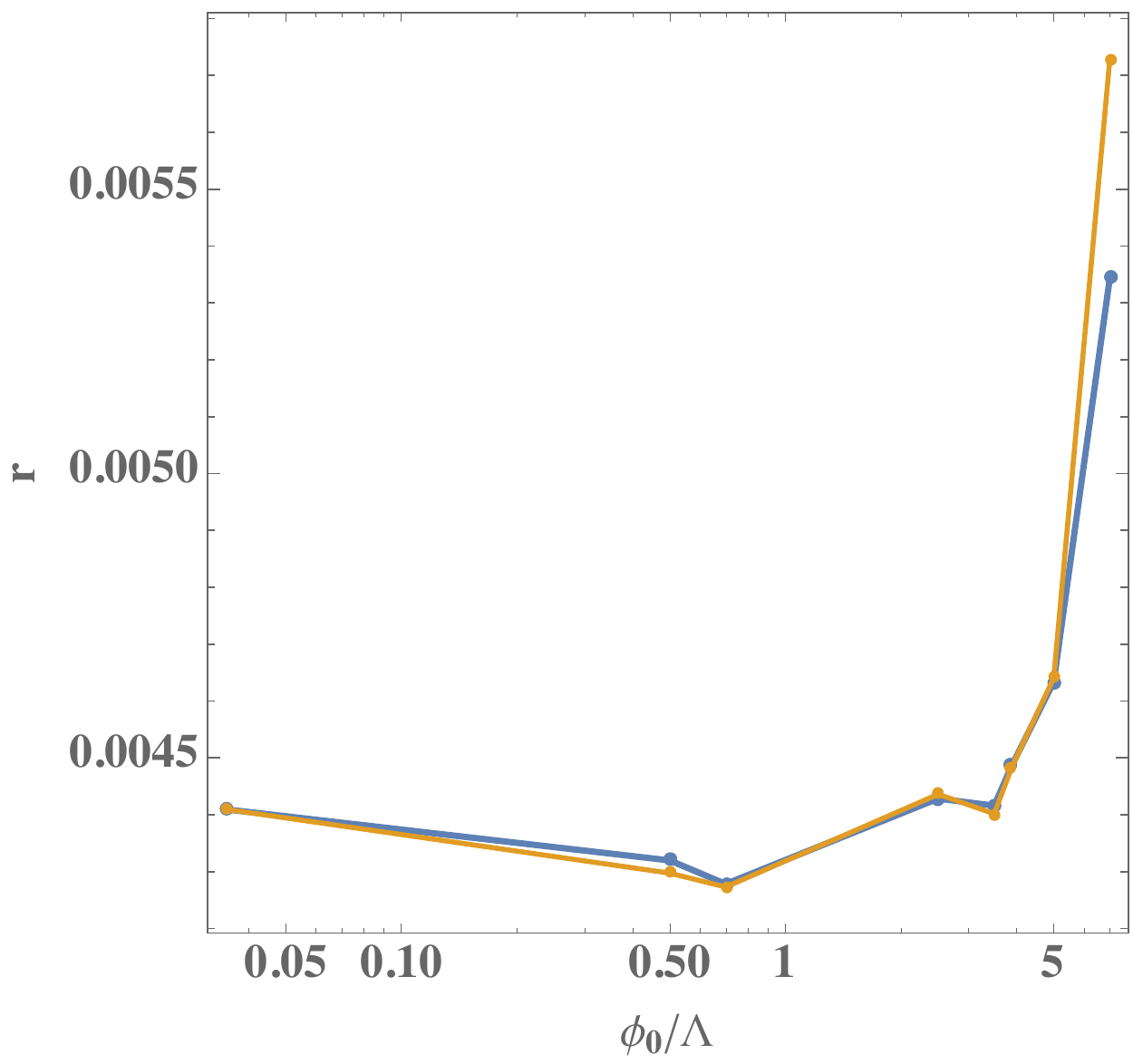}
	\end{minipage}
	\hspace{0.5cm}
	\begin{minipage}[b]{0.3\linewidth}
	\centering
	\includegraphics[width=\textwidth]{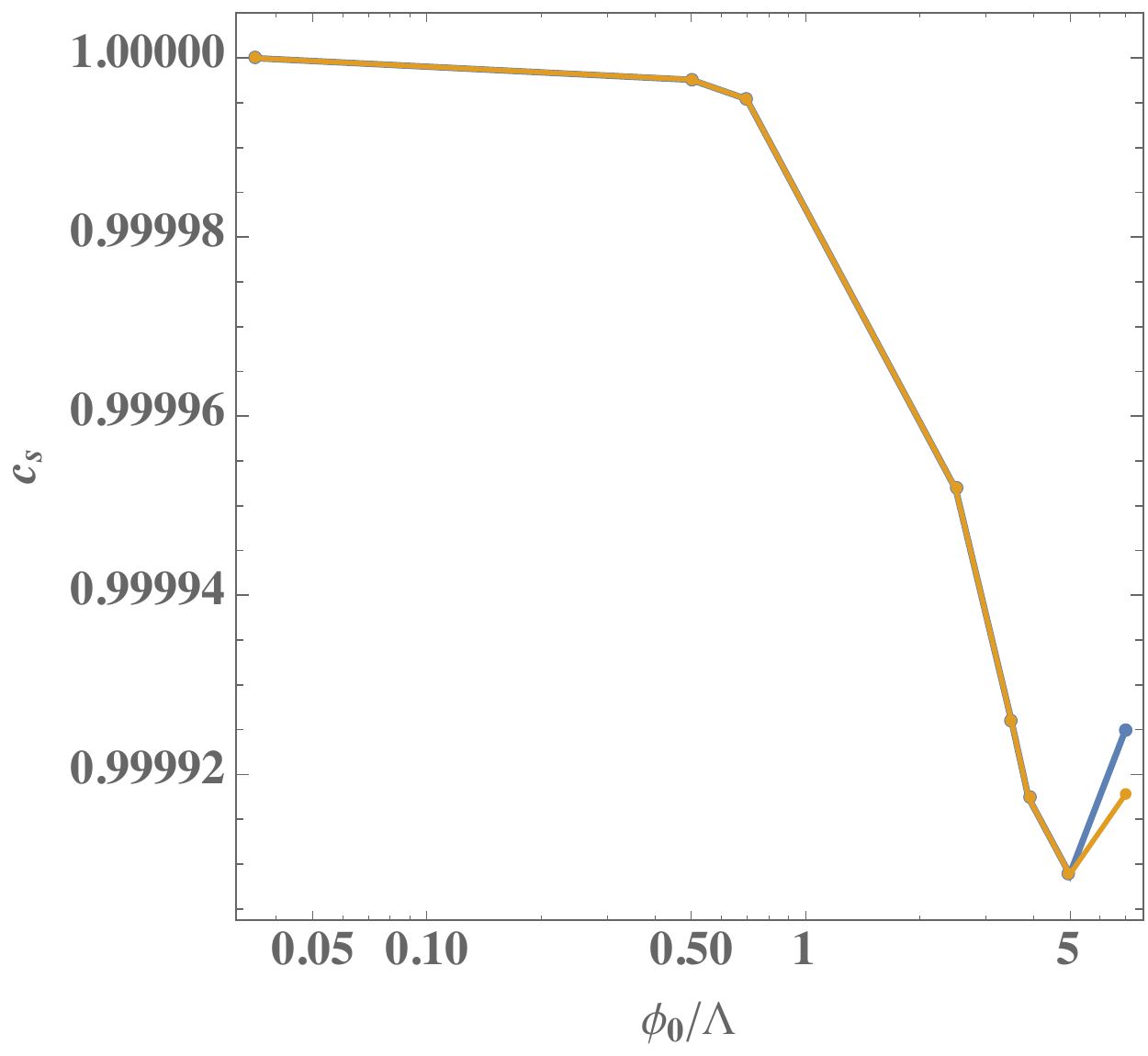}
    \end{minipage}
\caption{Observables for $f=e^{\phi_H/\Lambda}$ in the regime $m_L^2\ll g \phi_0^2$ and $ \frac{g \phi_L^2}{m_H^2}\gg1$. }
	\label{fig:ExpFlat}
\end{figure}

\item $\frac{g \phi_L^2}{m_H^2}\ll1$ \\

In this regime the potential takes the form
\be
V_{eff}(\Phi)=\frac{\mu^2}{2}\Phi^2\left(1- \frac{\mu^2 \Phi^2}{ m_H^2 \phi_0^2}+\frac{ 2 \mu^2 \Phi^2}{3 m_H^2 \phi_0^2}\frac{\phi_0}{\Lambda}\right)
\label{eq:Vexp2}
\ee
at leading order, where the effective mass parameter is $\mu^2\equiv g \phi_0^2 e^{-2 \phi_0/\Lambda}$. The speed of sound can schematically be written as 
\be
c_s^2-1=\sum_{n\ge1} \epsilon^n\left(\frac{\phi_0}{\Lambda}\right)^{2n} \left(\frac{\mu^2 \Phi^2}{\phi_0^2 m_H^2}\right)^n\ .
\ee
Noting that $\frac{g \phi_L^2}{m_H^2}\ll1\Rightarrow \frac{\mu^2 \Phi^2}{ m_H^2 \phi_0^2}\ll1$, decrease in $c_s$ can only be obtained in the large $\phi_0/\Lambda$ limit. In such a regime one observes that the contribution from the kinetic mixing to $V$, last term in Eq. \eqref{eq:Vexp2}, induces a steepening of the potential, that will ultimately dominate the scaling of the inflationary observables with $\phi_0/\Lambda$ as depicted in Fig. \ref{fig:ExpNonFlat}.

\end{itemize}

\section{Stringy musings}\label{sec:strings}

Axion monodromy inflation in string theory uses an axion field descending from the $p$-form gauge potentials of string theory to drive the inflationary dynamics. The scalar potential for the axion arises from fluxes or branes which provide an energy density for the axion displaying monodromy as a function of the axion traversing multiple fundamental periodicity domains. As compactified string theory generically contains several moduli scalar fields at mass scales below the KK scale, the axion monodromy potential experiences backreaction effects. These arise from the moduli dynamically adjusting to the presence of the potential energy stored in the inflaton axion and/or direct moduli-axion couplings in the kinetic terms or the scalar potential of the effective action.

A lot of the dynamics can be captured qualitatively by considering the simplest such class of systems, where the inflaton-axion couples to the lightest of the moduli, which produces a 2-field system. For many of the model classes of axion monodromy inflation studied in the literature, the 4D scalar field effective action of this 2-field system takes the form~\cite{Dong:2010in,McAllister:2014mpa}
\be
{\cal L}={\cal L}_K(\phi,\dot \phi,\dot b) - V(L,b)
\ee
with the scalar potential
\be
V(\phi,b)=\frac{C_1}{\phi^{q_1}}+C_2\,(\mu^2+b^2)\,\phi^{q_2}\quad.
\ee
The kinetic part of the action often takes two forms, depending on whether the lightest relevant modulus $L$ involved is one of the so-called 'bulk' volume or shape moduli, or whether it denotes a more 'local' deformation parameter taking only moderately large values. For a Calabi-Yau  (CY) compactification for instance, the 'bulk' volume or shape moduli describe the large overall volume and large complex structure directions of moduli space. 
\begin{itemize}
\item i) If the relevant lightest modulus is of 'bulk' type, then the kinetic part of the 4D effective action takes the form
\be
{\cal L}_K=\frac{k_1}{\phi^2}\dot \phi^2+\frac{k_2}{\phi^4}\dot b^2\quad.
\ee
An example here from type IIB flux compactifications would be e.g. $\phi\sim L$ with $L$ being the 'bulk' volume modulus of the internal space.
\item ii) For the non-bulk case the kinetic part often takes the form
\be
{\cal L}_K=K_1\phi^{p_1}\dot \phi^2+K_2\phi^{p_2}\dot b^2
\ee
with $p_1,p_2$ being positive ${\cal O}(1)$ constants and $K_1,K_2$ effective constants who are dominantly functions of the bulk moduli which in this case are assumed to be heavier than $\phi$ and thus effectively frozen. An example here are the blow-up volume moduli $v^i$of type IIB CY flux compactifications, where we would have $\phi\sim v$, and  $p_1,p_2=1$ while $K_1,K_2\sim 1/{\cal V}$.
\end{itemize}
From this structure it is immediately clear, that setups with kinetic functions of type ii) encompass the polynomial kinetic mixing examples of section~\ref{sec:KineticMix}, while all setups of type i) necessarily produce exponential kinetic mixing.

Finally, we look at the structure of the modulus-axion potential above. At the axion minimum $b=0$ the modulus $\phi$ acquires a minimum $\phi_0=\phi_0(\mu^2C_2/C_1,q_1/q_2)$. Hence, in the vicinity of this minimum the scalar potential has an expansion
\be
V(\phi,b)=m_0^2(\phi-\phi_0)^2+m_b^2\,b^2
\ee
where $m_0^2=m_0^2(\mu^2C_2/C_1,q_1/q_2)$ and $m_b^2=C_2\phi^{q_2}$. This potential is the starting point for the discussion in section~\ref{sec:PotKinMix}.
The string theory setups for axion monodromy inflation captured by the above EFT hence reduce in the appropriate limits to the toy models discussed in the previous sections.

\section{Summary}\label{sec:concl}

Inspired by the generic features of UV constructions of inflation, in particular by the presence of a kinetic coupling between the inflaton and heavier degrees of freedom, in this paper we developed EFT techniques to systematically study the observable signatures of such models. 

In the first part of the paper we explicitly demonstrated how theories with kinetic coupling between the inflaton and a heavy field can be equivalently described by two distinct EFTs. One may study the background at the two field level and then use an EFT for the perturbations, as done in e.g. \cite{Tolley:2009fg} and \cite{Achucarro:2010da}. Alternatively one may integrate out the heavy degree of freedom at the level of the background, thereby obtaining a background EFT with HD interactions. We have developed a recursive method that allows us to demonstrate that these two EFTs are equivalent and yield results compatible with the computationally costlier two field computation. Our results hold in the presence of both kinetic and potential mixing. In both cases the end result is a reduced propagation speed for the scalar perturbations, a parameter  which has an important impact on the inflationary observables and one for which current observations only place a weak lower bound.

In the second part of this work we presented specific examples to illustrate the effects of the kinetic (and potential) interaction between heavy and light fields. If the mixing happens exclusively via the kinetic term, the main effect one finds is a reduction of the speed of sound that entails a reduction of the tensor-to-scalar ratio. One can in fact decrease $r$ significantly while having little or no effect on the tilt of the scalar power spectrum. Specific examples of this behaviour have been previously reported in e.g. \cite{Achucarro:2015rfa}. We stress that this can be done within the regime of validity of the EFTs, with the heavy field above the Hubble scale. This mechanism can be used to decrease the current tension between chaotic inflation models and observational data. It is worth pointing out that there are limits to how much $r$ can be reduced  \cite{Leblond:2008gg,Stein:2016jja} and that future CMB polarization observations \cite{Kallosh:2019eeu} will be able to place more stringent bounds on it, so chaotic inflation models can still be ruled out.  If on top of the kinetic interaction one also considers mixing in the potential the situation changes drastically due to the non-trivial canonical normalisation of the inflaton. In such more complex cases, the behaviour of the observables is more model and parameter dependent and one can find instances where $c_s$ and $r$ can be simultaneously reduced and others where a reduction of $c_s$ is attainable but comes associated with an increase of $r$. Either way the present work confirms that the observable signatures of simple looking two field models can be much richer than what one would find in the simplest two derivative EFT analysis. From the observational side, a significant reduction of the upper bound on $r$ or of the upper bounds on $f_{NL}$ would be most welcome in order to further constrain this class of UV-inspired models of inflation.

\section*{Acknowledgments}
We would like to thank Y. Welling for many 
very useful discussions.   AW is supported by the ERC Consolidator Grant STRINGFLATION under the HORIZON 2020 grant agreement no. 647995, as well as by the Deutsche Forschungsgemeinschaft (DFG, German Research Foundation) under Germany's Excellence Strategy -- EXC 2121 ``Quantum Universe'' -- 390833306.


\begin{thebibliography}{99}
\bibitem{Akrami:2018odb}
  Y.~Akrami {\it et al.} [Planck Collaboration],
  ``Planck 2018 results. X. Constraints on inflation,''
  arXiv:1807.06211 [astro-ph.CO].


\bibitem{Akrami:2019izv}
  Y.~Akrami {\it et al.} [Planck Collaboration],
  ``Planck 2018 results. IX. Constraints on primordial non-Gaussianity,''
  arXiv:1905.05697 [astro-ph.CO].


\bibitem{Chen:2006nt} 
  X.~Chen, M.~x.~Huang, S.~Kachru and G.~Shiu,
  ``Observational signatures and non-Gaussianities of general single field inflation,''
  JCAP {\bf 0701}, 002 (2007)
  doi:10.1088/1475-7516/2007/01/002
  [hep-th/0605045].
  
\bibitem{Arkani-Hamed:2015bza} 
  N.~Arkani-Hamed and J.~Maldacena,
  ``Cosmological Collider Physics,''
  arXiv:1503.08043 [hep-th].

\bibitem{Arkani-Hamed:2018kmz} 
  N.~Arkani-Hamed, D.~Baumann, H.~Lee and G.~L.~Pimentel,
  ``The Cosmological Bootstrap: Inflationary Correlators from Symmetries and Singularities,''
  arXiv:1811.00024 [hep-th].

\bibitem{Alishahiha:2004eh}
  M.~Alishahiha, E.~Silverstein and D.~Tong,
  ``DBI in the sky,''
  Phys.\ Rev.\ D {\bf 70} (2004) 123505
  doi:10.1103/PhysRevD.70.123505
  [hep-th/0404084].


\bibitem{Cheung:2007st}
  C.~Cheung, P.~Creminelli, A.~L.~Fitzpatrick, J.~Kaplan and L.~Senatore,
  ``The Effective Field Theory of Inflation,''
  JHEP {\bf 0803} (2008) 014
  doi:10.1088/1126-6708/2008/03/014
  [arXiv:0709.0293 [hep-th]].


\bibitem{Baumann:2011su}
  D.~Baumann and D.~Green,
  ``Equilateral Non-Gaussianity and New Physics on the Horizon,''
  JCAP {\bf 1109} (2011) 014
  doi:10.1088/1475-7516/2011/09/014
  [arXiv:1102.5343 [hep-th]].


\bibitem{Leblond:2008gg}
  L.~Leblond and S.~Shandera,
  ``Simple Bounds from the Perturbative Regime of Inflation,''
  JCAP {\bf 0808} (2008) 007
  doi:10.1088/1475-7516/2008/08/007
  [arXiv:0802.2290 [hep-th]].


\bibitem{Dong:2010in}
  X.~Dong, B.~Horn, E.~Silverstein and A.~Westphal,
  ``Simple exercises to flatten your potential,''
  Phys.\ Rev.\ D {\bf 84} (2011) 026011
  doi:10.1103/PhysRevD.84.026011
  [arXiv:1011.4521 [hep-th]].


\bibitem{McAllister:2014mpa}
  L.~McAllister, E.~Silverstein, A.~Westphal and T.~Wrase,
  ``The Powers of Monodromy,''
  JHEP {\bf 1409} (2014) 123
  doi:10.1007/JHEP09(2014)123
  [arXiv:1405.3652 [hep-th]].

\bibitem{Kaloper:2011jz} 
  N.~Kaloper, A.~Lawrence and L.~Sorbo,
  ``An Ignoble Approach to Large Field Inflation,''
  JCAP {\bf 1103}, 023 (2011)
  doi:10.1088/1475-7516/2011/03/023
  [arXiv:1101.0026 [hep-th]].

\bibitem{DAmico:2017cda} 
  G.~D'Amico, N.~Kaloper and A.~Lawrence,
  ``Monodromy Inflation in the Strong Coupling Regime of the Effective Field Theory,''
  Phys.\ Rev.\ Lett.\  {\bf 121}, no. 9, 091301 (2018)
  doi:10.1103/PhysRevLett.121.091301
  [arXiv:1709.07014 [hep-th]].

\bibitem{Tolley:2009fg}
  A.~J.~Tolley and M.~Wyman,
  ``The Gelaton Scenario: Equilateral non-Gaussianity from multi-field dynamics,''
  Phys.\ Rev.\ D {\bf 81} (2010) 043502
  doi:10.1103/PhysRevD.81.043502
  [arXiv:0910.1853 [hep-th]].





\bibitem{Achucarro:2015rfa}
  A.~Achucarro, V.~Atal and Y.~Welling,
  ``On the viability of $m^2\phi^2$ and natural inflation,''
  JCAP {\bf 1507} (2015) 008
  doi:10.1088/1475-7516/2015/07/008
  [arXiv:1503.07486 [astro-ph.CO]].


\bibitem{Achucarro:2010da}
  A.~Achucarro, J.~O.~Gong, S.~Hardeman, G.~A.~Palma and S.~P.~Patil,
  ``Features of heavy physics in the CMB power spectrum,''
  JCAP {\bf 1101} (2011) 030
  doi:10.1088/1475-7516/2011/01/030
  [arXiv:1010.3693 [hep-ph]].


\bibitem{Sasaki:1995aw}
  M.~Sasaki and E.~D.~Stewart,
  ``A General analytic formula for the spectral index of the density perturbations produced during inflation,''
  Prog.\ Theor.\ Phys.\  {\bf 95} (1996) 71
  doi:10.1143/PTP.95.71
  [astro-ph/9507001].


\bibitem{Gordon:2000hv}
  C.~Gordon, D.~Wands, B.~A.~Bassett and R.~Maartens,
  ``Adiabatic and entropy perturbations from inflation,''
  Phys.\ Rev.\ D {\bf 63} (2001) 023506
  doi:10.1103/PhysRevD.63.023506
  [astro-ph/0009131].


\bibitem{GrootNibbelink:2001qt}
  S.~Groot Nibbelink and B.~J.~W.~van Tent,
  ``Scalar perturbations during multiple field slow-roll inflation,''
  Class.\ Quant.\ Grav.\  {\bf 19} (2002) 613
  doi:10.1088/0264-9381/19/4/302
  [hep-ph/0107272].


\bibitem{Burgess:2012dz}
  C.~P.~Burgess, M.~W.~Horbatsch and S.~P.~Patil,
  ``Inflating in a Trough: Single-Field Effective Theory from Multiple-Field Curved Valleys,''
  JHEP {\bf 1301} (2013) 133
  doi:10.1007/JHEP01(2013)133
  [arXiv:1209.5701 [hep-th]].


\bibitem{Gong:2014rna}
  J.~O.~Gong, M.~S.~Seo and S.~Sypsas,
  ``Higher derivatives and power spectrum in effective single field inflation,''
  JCAP {\bf 1503} (2015) no.03,  009
  doi:10.1088/1475-7516/2015/03/009
  [arXiv:1407.8268 [hep-th]].


\bibitem{Weinberg:2008hq}
  S.~Weinberg,
  ``Effective Field Theory for Inflation,''
  Phys.\ Rev.\ D {\bf 77} (2008) 123541
  doi:10.1103/PhysRevD.77.123541
  [arXiv:0804.4291 [hep-th]].


\bibitem{ArmendarizPicon:1999rj}
  C.~Armendariz-Picon, T.~Damour and V.~F.~Mukhanov,
  ``k - inflation,''
  Phys.\ Lett.\ B {\bf 458} (1999) 209
  doi:10.1016/S0370-2693(99)00603-6
  [hep-th/9904075].


\bibitem{Garriga:1999vw}
  J.~Garriga and V.~F.~Mukhanov,
  ``Perturbations in k-inflation,''
  Phys.\ Lett.\ B {\bf 458} (1999) 219
  doi:10.1016/S0370-2693(99)00602-4
  [hep-th/9904176].


\bibitem{Cremonini:2010ua}
  S.~Cremonini, Z.~Lalak and K.~Turzynski,
  ``Strongly Coupled Perturbations in Two-Field Inflationary Models,''
  JCAP {\bf 1103} (2011) 016
  doi:10.1088/1475-7516/2011/03/016
  [arXiv:1010.3021 [hep-th]].


\bibitem{Achucarro:2010jv}
  A.~Achucarro, J.~O.~Gong, S.~Hardeman, G.~A.~Palma and S.~P.~Patil,
  ``Mass hierarchies and non-decoupling in multi-scalar field dynamics,''
  Phys.\ Rev.\ D {\bf 84} (2011) 043502
  doi:10.1103/PhysRevD.84.043502
  [arXiv:1005.3848 [hep-th]].


\bibitem{Sasaki:1986hm}
  M.~Sasaki,
  ``Large Scale Quantum Fluctuations in the Inflationary Universe,''
  Prog.\ Theor.\ Phys.\  {\bf 76} (1986) 1036.
  doi:10.1143/PTP.76.1036


\bibitem{Mukhanov:1988jd}
  V.~F.~Mukhanov,
  ``Quantum Theory of Gauge Invariant Cosmological Perturbations,''
  Sov.\ Phys.\ JETP {\bf 67} (1988) 1297
   [Zh.\ Eksp.\ Teor.\ Fiz.\  {\bf 94N7} (1988) 1].


\bibitem{Avgoustidis:2011em}
  A.~Avgoustidis, S.~Cremonini, A.~C.~Davis, R.~H.~Ribeiro, K.~Turzynski and S.~Watson,
  ``The Importance of Slow-roll Corrections During Multi-field Inflation,''
  JCAP {\bf 1202} (2012) 038
  doi:10.1088/1475-7516/2012/02/038
  [arXiv:1110.4081 [astro-ph.CO]].


\bibitem{Cespedes:2012hu}
  S.~Cespedes, V.~Atal and G.~A.~Palma,
  ``On the importance of heavy fields during inflation,''
  JCAP {\bf 1205} (2012) 008
  doi:10.1088/1475-7516/2012/05/008
  [arXiv:1201.4848 [hep-th]].


\bibitem{Achucarro:2012yr}
  A.~Achucarro, V.~Atal, S.~Cespedes, J.~O.~Gong, G.~A.~Palma and S.~P.~Patil,
  ``Heavy fields, reduced speeds of sound and decoupling during inflation,''
  Phys.\ Rev.\ D {\bf 86} (2012) 121301
  doi:10.1103/PhysRevD.86.121301
  [arXiv:1205.0710 [hep-th]].


\bibitem{Stein:2016jja}
  N.~K.~Stein and W.~H.~Kinney,
  ``Planck Limits on Non-canonical Generalizations of Large-field Inflation Models,''
  JCAP {\bf 1704} (2017) no.04,  006
  doi:10.1088/1475-7516/2017/04/006
  [arXiv:1609.08959 [astro-ph.CO]].


\bibitem{Freese:1990rb}
  K.~Freese, J.~A.~Frieman and A.~V.~Olinto,
  ``Natural inflation with pseudo - Nambu-Goldstone bosons,''
  Phys.\ Rev.\ Lett.\  {\bf 65} (1990) 3233.
  doi:10.1103/PhysRevLett.65.3233


\bibitem{Kallosh:2019eeu}
  R.~Kallosh and A.~Linde,
  ``B-mode Targets,''
  arXiv:1906.04729 [astro-ph.CO].

\end{thebibliography}
\end{document}